\documentclass[submission, Phys]{SciPost}

\usepackage{graphicx}
\usepackage{bm}
\usepackage{dsfont}
\usepackage[dvipsnames]{xcolor}
\usepackage{pstricks}
\usepackage{verbatim}
\usepackage{units}
\usepackage{multirow}
\usepackage{enumitem}
\usepackage{mathrsfs}
\usepackage{leftidx}
\usepackage{xspace}
\usepackage{braket}
\usepackage{bbm}
\usepackage{cancel}
\usepackage{amsfonts,amssymb,amsmath}
\usepackage[normalem]{ulem}
\usepackage{xr}

\DeclareMathAlphabet{\mathpgoth}{OT1}{pgoth}{m}{n}

\hypersetup{colorlinks=true,linktoc=all,linkcolor=blue,breaklinks=true,citecolor=blue,urlcolor=blue}
\usepackage[english]{babel}
\usepackage{cleveref}
\usepackage{comment}

\newcommand{\ketbra}[2]{\ket{#1}\!\!\bra{#2}}
\newcommand{\proj}[1]{\ketbra{#1}{#1}}

\newcommand{\Tr}{\text{Tr}}

\renewcommand{\S}{\mathcal{S}}
\newcommand{\Shannon}{S_\text{Sh}}

\newcommand{\Ex}[1]{\mathbb{E}\!\left[#1\right]}
\newcommand{\Var}[1]{\text{Var}\left[#1\right]}
\newcommand{\QQuad}{\qquad\qquad}
\newcommand{\cut}{\!\!\!\!}

\newcommand\smallO{
  \mathchoice
    {{\scriptstyle\mathcal{O}}}
    {{\scriptstyle\mathcal{O}}}
    {{\scriptscriptstyle\mathcal{O}}}
    {\scalebox{.7}{$\scriptscriptstyle\mathcal{O}$}}
  }

\numberwithin{equation}{section}

\begin{document}

\begin{center}{\Large \textbf{
Stochastic entropy production in scattering theory
}}\end{center}

\begin{center}
Ludovico Tesser\textsuperscript{1*},
Henning Kirchberg\textsuperscript{1},
Matteo Acciai\textsuperscript{2},
Janine Splettstoesser\textsuperscript{1}
\end{center}

\begin{center}
{\bf 1} Department of Microtechnology and Nanoscience (MC2), Chalmers University of Technology, Kemivägen 9, 41296 Göteborg, Sweden
\\
{\bf 2} The Abdus Salam International Centre for Theoretical Physics (ICTP), Strada Costiera 11, 34151 Trieste, Italy
\\
* tesser@chalmers.se
\end{center}

\begin{center}
\today
\end{center}


\section*{Abstract}
{\bf
We formulate a stochastic description of entropy production in scattering theory for coherent transport.
We distinguish between the information entropy change due to partial knowledge of the leads' state and the thermodynamic entropy change due to the equilibration of each lead with its bath.
By employing a two-point measurement scheme, we access the stochastic entropy production at these different stages of the process, as well as the statistics of generic transport quantities.
When restricted to particle or energy transport, our approach reproduces the Landauer-B\"uttiker formulas.
The possibility to consider more general quantities such as the entropy currents and their fluctuations, provides a systematic connection between stochastic thermodynamics and coherent transport.
}

\vspace{10pt}
\noindent\rule{\textwidth}{1pt}
\tableofcontents\thispagestyle{fancy}
\noindent\rule{\textwidth}{1pt}
\vspace{10pt}

\section{Introduction}

Entropy production is a key quantity in quantum stochastic thermodynamics that relates thermodynamics and information, and quantifies irreversibility even at the level of single realizations~\cite{Strasberg2022Jan, Campbell2026Jan}.
In systems weakly coupled to their baths, the thermodynamic framework is well understood, and allows to characterize not only average quantities, but also fluctuations~\cite{Landi2024Apr}. In particular, the development of so-called uncertainty relations~\cite{Barato2015Apr, DiTerlizzi2018Dec, Hasegawa2020Jul, Prech2025Jan}, valid for a variety of currents and systems, connects the dissipation incurred by a system and the current precision that can be achieved.

By contrast, it is a significant challenge to establish such a thermodynamic framework for systems that are strongly coupled to their baths, especially because the distinction between system and bath is not clear-cut.
A breadth of methods to tackle such strongly-coupled systems is currently being used~\cite{ThermoQuantumRegime2018}, including the reaction coordinate mapping~\cite{Strasberg2016Jul, Strasberg2018May,Landi2022Dec}, the mesoscopic lead formulation~\cite{Lacerda2023May, Brenes2023Aug, Bettmann2025Jul}, as well as nonequilibrium Green's functions~\cite{esp2015, ses2021feb, zhou2024feb}.
Coherent quantum transport provides an instance of such strongly-coupled setups where a central system or conductor is (strongly) coupled to leads. A powerful method to describe these systems in the limit of weak many-body interactions is scattering theory. These coherent conductors are of high interest in the context of thermodynamics~\cite{Benenti2017Jun,Campbell2026Jan}. For example, they exhibit violations of the uncertainty relations initially developed in weak coupling~\cite{Saryal2019Oct, Brandner2018Mar,Brandner2025Jul,Ehrlich2021Jul,Palmqvist2025Jul, Palmqvist2025Oct,Blasi2025May}, meaning that higher current precision can be achieved beyond weak coupling. However, understanding how and to what extent the uncertainty relations can be recovered in such systems would require a stochastic description of currents and entropy production, which is lacking in scattering theory. 
This central gap, namely the missing stochastic description of thermodynamically relevant quantities in scattering theory, is addressed in this work.

\begin{figure}[hb!]
    \centering
    \includegraphics[scale=1.2]{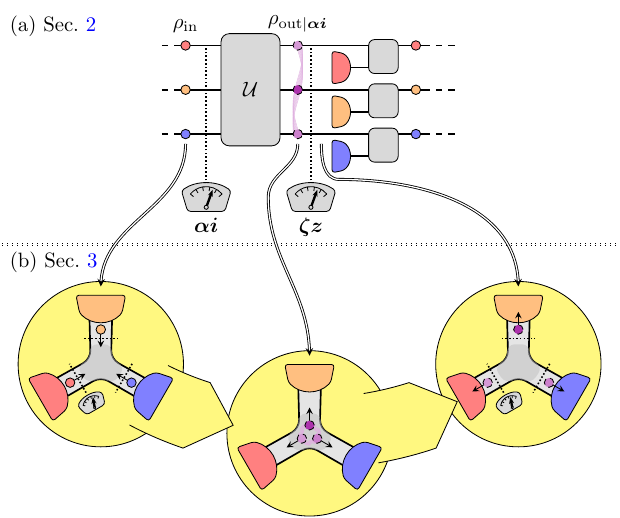}
    \caption{(a) Schematic representation of the process considered. The input state $\rho_\text{in}$ is separable and undergoes a (non-perturbing) measurement with outcome $\boldsymbol{\alpha i}$. After a unitary evolution, the (conditioned) output state $\rho_{\text{out}|\boldsymbol{\alpha i}}$ is measured again. Then, the system's state is reset by making each subsystem interact with a bath.
    In Sec.~\ref{sec:info-thermo-entropy}, we focus on the entropy change at different stages of the process.
    (b) Scattering process of Sec.~\ref{sec:Scattering}. The input state is initially separable and undergoes a non-perturbing measurement. The scattering process implements a unitary transformation on the state, creating correlations between the leads. A measurement on the output state destroys the correlations, and finally the state on each lead thermalizes with the corresponding bath.
    }
    \label{fig}
\end{figure}

The key thermodynamic quantity that we consider here is entropy; however our approach can straightforwardly be extended to other thermodynamically relevant quantities that rely on the knowledge of the system state. We study a generic quantum conductor that is possibly subject to periodic time-dependent driving and coupled to multiple terminals via multi-channel leads, see Fig.~\ref{fig}. We build upon prior works that investigated the \textit{average} entropy production using the so-called inside-outside duality~\cite{bru2018Mar, zhou2024feb}, where the entropy production in the scattering region is inferred from the one in the leads.
Here, we also focus on the leads' state, but we first employ elements from the repeated interaction framework~\cite{Strasberg2017Apr, Strasberg2019Oct, Bettmann2023Apr} to set the stage and define thermodynamic quantities following Ref.~\cite{Esposito2010Jan}.
Concretely, this means that we consider the transport process at different stages, separating it into the unitary process of the scattering event (described by the scattering matrix), followed by an interaction with the bath.
The separation between the unitary transformation on the system and the subsystem-bath couplings allows us to distinguish between the \textit{information} entropy change in the leads and the \textit{thermodynamic} entropy change in the corresponding bath.
To access fluctuations~\cite{Esposito2009Dec, Landi2021Sep} and define stochastic quantities---in particular the stochastic entropy production---we introduce a two-point measurement scheme: A separable initial state is measured; then it undergoes a unitary transformation (the scattering process), and is subsequently measured again.
Afterwards, each subsystem is reset to its initial state by coupling with a corresponding bath. The outcomes of the measurement process can then directly be connected to measurable currents into the terminals (baths)~\cite{Tesser2025-thesis}. This procedure not only recovers the well-known results for particle and energy currents and their fluctuations from Landauer-B\"uttiker scattering theory, but it also provides a rigorous derivations of entropy currents~\cite{deg2020} and in particular their fluctuations~\cite{Acciai2024Feb}.
Importantly, this treatment includes the case of baths described by \textit{nonthermal} occupations. Such nonthermal baths can be used as resources, for instance, to refrigerate while providing neither energy nor particles to the working substance \textit{on average}~\cite{Sanchez2019Nov, deg2020}, while crucially requiring input fluctuations~\cite{Acciai2024Feb, Monsel2025Oct}.
We expect that the connection between stochastic thermodynamics and quantum transport provided in this work will be beneficial to analyze thermodynamically relevant quantities beyond stochastic entropy in quantum transport in the future.

The paper is structured as follows. In Sec.~\ref{sec:info-thermo-entropy}, we start with a general discussion of a process involving measurements, unitary evolution, and dissipation, as depicted in Fig.~\ref{fig}(a). Next, in Sec.~\ref{sec:Scattering}, we apply this framework to a single coherent scattering process, where the subsystems studied in Sec.~\ref{sec:info-thermo-entropy} represent the leads connected to a central scattering region in quantum transport. Here, we derive the marginal states of the leads and the corresponding entropy changes. We also introduce the two-point measurement scheme to access fluctuations. Finally, in Sec.~\ref{sec:scattering_theory}, we connect the single-event statistics to generic average currents, such as entropy currents, and their zero-frequency noise in scattering theory, recovering also expected results such as the Landauer-B\"uttiker formula for charge currents and noise. The conclusions are drawn in Sec.~\ref{sec:conclusions}, where we outline the implications and future perspectives of this work.
Technical details of the derivations are provided in four appendices.


\section{Information and thermodynamic entropy change}\label{sec:info-thermo-entropy}

We consider a generic multipartite quantum system that undergoes a unitary transformation mapping the input state $\rho_\text{in}$ into the output state $\rho_\text{out} = \mathcal{U}\rho_\text{in} = U\rho_\text{in} U^\dagger$, as sketched in Fig.~\ref{fig}(a).
Such a transformation could for example be used to prepare a ``useful" state to be employed subsequently for a ``task", e.g. energy harvesting. This transformation can generate correlations between the subsystems, while preserving the von Neumann entropy $S[\rho]:=-\Tr\{\rho\log\rho\}$ of the full system
\begin{equation}\label{eq:von_Neumann_entropy_conserved}
    S[\rho_{\mathrm{out}}] = S[\rho_{\mathrm{in}}].
\end{equation}
Importantly, signatures of these correlations are found in the change of the von Neumann entropy of the \textit{reduced states} of the subsystems or of a projected state due to a measurement, as discussed in Sec.~\ref{sec:informational-entropy}.

Subsequently, each subsystem $\alpha$ is coupled to a corresponding bath $\alpha$, which is initially uncorrelated to the system and prepared in the state  $\rho_\alpha^\text{B}(0)$.
The subsystem and the bath interact until the subsystem eventually equilibrates with the (much larger) bath, corresponding to a ``reset" of the state. In this process, each subsystem exchanges both particles and energy with the bath, generating a thermodynamic entropy production, discussed in Sec.~\ref{sec:thermodynamic-entropy}.

\subsection{Change of information entropy}\label{sec:informational-entropy}
In the initial unitary transformation that the system undergoes, the different subsystems are coupled together, generating a state that spans over the joint Hilbert space $\mathcal{H}_1\otimes\cdots\otimes \mathcal{H}_L$. However, when the different subsystems are accessed locally, only the information contained in the reduced density matrix is accessible.

This is for instance the case when measurements in one of the subsystems (later in Sec.~\ref{sec:Scattering} the leads of the coherent conductor) are considered. Focusing without loss of generality on the first subsystem, $\alpha=1$, its input/output density matrices are obtained by tracing out all other subsystems (denoted by $\Tr_{\bar{1}}$), captured by $\rho_{1,\text{in/out}} = \Tr_{\bar{1}}{\rho_\text{in/out}}$.
Then, we refer to the change in the von Neumann entropy between these two states, namely
\begin{equation}\label{eq:info_entropy}
    \Delta S_1^\mathrm{info} = S[\rho_{1,\mathrm{out}}]-S[\rho_{1,\mathrm{in}}],
\end{equation}
as change of information entropy, which depends on the correlations generated by the unitary transformation.
For instance, when the system is initially prepared in a separable state, i.e. $\rho_\text{in}= \bigotimes_\alpha \rho_{\alpha,\text{in}}$, the sum of information entropy change in all subsystems reads
\begin{equation}
\begin{split}
    \sum_\alpha \Delta S_\alpha^\mathrm{info}& := S\left[\bigotimes_\alpha \rho_{\alpha,\text{out}}\right] -S\left[\bigotimes_\alpha \rho_{\alpha,\text{in}}\right] = S\left[\bigotimes_\alpha \rho_{\alpha,\text{out}}\right] -S\left[\rho_{\text{out}}\right]\\
    &=D\left[\rho_{\text{out}}\big|\big|\bigotimes_\alpha \rho_{\alpha,\text{out}}\right] \geq 0,
\end{split}
\end{equation}
where $D[\rho||\sigma]:=\Tr\{\rho\log\rho - \rho\log\sigma\}$ is the relative entropy.
Here, we used the invariance of the von Neumann entropy under the unitary transformation mapping $\rho_\text{in}$ into $\rho_\text{out}$ [Eq.~\eqref{eq:von_Neumann_entropy_conserved}], and we recognized the relative entropy between the output state and the tensor product over the subsystems of its marginal states.
The non-negativity of the relative entropy therefore tells us that the presence of entanglement between the subsystems is reflected in a positive change of the total information entropy.

Concretely, to evaluate the information entropy Eq.~\eqref{eq:info_entropy} one needs to access the full reduced density matrix of the subsystem. In practice, this requires quantum state tomography on the subsystems~\cite{paris2004quantum, Cramer2010Dec, Huang2020Oct}, which is particularly challenging in electronic transport~\cite{Jullien2014Oct, Bisognin2019Jul, Bourgeois2026Jan}.
A less demanding (though not informationally complete) way to characterize the state of the system is by measuring an observable of the system.
Here, we focus on the case in which the measurement is characterized by rank-1 projectors $\{\hat{\Pi}_m\}$, such that $\hat{\Pi}_m\hat{\Pi}_{m'} = \delta_{m,m'}\hat{\Pi}_m$ and $\sum_m \hat{\Pi}_m = \mathbbm{1}$.
Then, the (average) action of the measurement on the state is given by
\begin{equation}\label{eq:avg_measurement_action}
  \mathcal{M} \rho := \sum_m \hat{\Pi}_m\rho \hat{\Pi}_m,
\end{equation}
which destroys the coherences in the measurement basis.
When this measurement is used to reconstruct input ($\rho_\text{in}$) and output ($\rho_\text{out}$) states \textit{separately} and \textit{independently}, one attributes the change of entropy
\begin{equation}\label{eq:measurement_entropy}
   \Delta S^\text{meas} := S[\mathcal{M}\rho_\text{out}] - S[\mathcal{M}\rho_\text{in}]
\end{equation}
to the process.
This is particularly useful when the initial state is not perturbed by the measurement, i.e. $\mathcal{M}\rho_\text{in}=\rho_\text{in}$. Then, $S[\mathcal{M}\rho_\text{in}] = S[\rho_\text{in}]$ and the measured entropy change satisfies
\begin{equation}\label{eq:meas_entropy}
 \Delta S^\text{meas}\geq S[\rho_\text{out}] -S[\rho_\text{in}]=0,  
\end{equation}
where we used that the measurements projectors have rank 1.
When the measurement is performed only locally on a subsystem, e.g. on subsystem 1, we replace the projectors $\{\hat{\Pi}_m\}$ with projectors acting only on the given subsystem $\{\hat{\Pi}_{1,m}\}$. We refer to the (average) action of the local measurement as $\mathcal{M}_1$, analogously to Eq.~\eqref{eq:avg_measurement_action}.
Then, in the same fashion as Eq.~\eqref{eq:measurement_entropy}, the local measured entropy change reads
\begin{equation}
   \Delta S^\text{meas}_1 := S[\mathcal{M}_1\rho_\text{1,out}] - S[\mathcal{M}_1\rho_\text{1,in}],
\end{equation}
and, when the input state is unperturbed by the measurement $\mathcal{M}_1\rho_\text{1,in} = \rho_\text{1,in}$, fulfills
\begin{equation}\label{eq:meas_entropy1}
       \Delta S^\text{meas}_1\geq S[\rho_\text{1,out}] - S[\rho_\text{1,in}] = \Delta S_1^\text{info}.
\end{equation}

In Sec.~\ref{sec:Scattering}, we apply this framework to describe the information entropy produced in a scattering event in coherent transport, as shown in Fig.~\ref{fig}(b).
There, the subsystems are the different leads connecting the respective baths to the central scattering region.
The baths prepare the initial state of the leads $\rho_{\alpha,\text{in}}$, while the scattering region implements the unitary transformation.

The difference between the entropy changes~(\ref{eq:von_Neumann_entropy_conserved}, \ref{eq:info_entropy}) of the (sub)system's state and the ones obtained by reconstructing the state through a measurement, Eqs.~(\ref{eq:meas_entropy}, \ref{eq:meas_entropy1}), is clarified in Sec.~\ref{subsec:2PM}.
There, we employ the two-point measurement scheme to access information and statistics of the system, allowing us to define the stochastic entropy change of single realizations of the scattering process.

\subsection{Change of thermodynamic entropy due to system-bath coupling}\label{sec:thermodynamic-entropy}

After the unitary transformation on the multipartite system, each subsystem is coupled (from time $0$ to $t$) to the corresponding bath in order to reset the state to $\rho_\mathrm{in}$.
We now focus on the entropy change due to the coupling between (without loss of generality) subsystem 1 and its bath, as indicated by the subscript ``1" in the following.
Following Ref.~\cite{Esposito2010Jan}, the thermodynamic entropy production due to the coupling of a quantum system to a bath fulfills the second law as derived from the relative entropy
\begin{equation}\label{eq:2nd-law-relative-entropy0}
\begin{split}
    &D\left[\rho_1^\mathrm{S+B}(t)||\rho_1^\mathrm{S}(t)\otimes\rho_1^\mathrm{B}(t)\right]  =  \mathrm{Tr}\left\{\rho_1^\mathrm{S+B}(0)\log \rho_1^\mathrm{S+B}(0)-\rho_1^\mathrm{S+B}(t)\log [\rho_1^\mathrm{S}(t)\otimes\rho_1^\mathrm{B}(t)]\right\}\\
    &=- \mathrm{Tr}\left\{\rho_1^\mathrm{S}(t)\log \rho_1^\mathrm{S}(t)-\rho_1^\mathrm{S}(0)\log \rho_1^\mathrm{S}(0)\right\} - \mathrm{Tr}\left\{\rho_1^\mathrm{B}(t)\log \rho_1^\mathrm{B}(t)-\rho_1^\mathrm{B}(0)\log \rho_1^\mathrm{B}(0)\right\}.\\
\end{split}
\end{equation}
In the first line we used that (minus) the von Neumann entropy of the joint subsystem and bath density matrix is invariant under the unitary evolution, allowing us to replace the time argument $t$ with the initial time $t=0$ as argument of $\rho_1^\mathrm{S+B}$.
In the second line, we assumed that the subsystem and the bath are initially uncorrelated, i.e. $\rho_1^\mathrm{S+B}(0)=\rho_1^\mathrm{S}(0)\otimes\rho_1^\mathrm{B}(0)$.

From Eq.~\eqref{eq:2nd-law-relative-entropy0}, we therefore decompose the relative entropy as the sum of the von Neumann entropy changes of the reduced subsystem and bath
\begin{equation}\label{eq:2nd-law-relative-entropy}
    D\left[\rho_1^\mathrm{S+B}(t)||\rho_1^\mathrm{S}(t)\otimes\rho_1^\mathrm{B}(t)\right] = \Delta S_{1}^\mathrm{S}+\Delta S_{1}^\mathrm{B}\geq 0.
\end{equation}
Crucially, the subsystem's state at the beginning of the system-bath coupling is given by the output state of the first unitary transformation, i.e.  $\rho_1^\text{S}(0) = \rho_{1,\text{out}}$.
Additionally, we let the subsystem equilibrate with the (much larger) bath, such that at the end of the system-bath coupling the reduced state of the subsystem coincides with the state prepared by the bath at the beginning of the whole process, i.e. $\rho_1^\text{S}(t) = \rho_{1,\text{in}}$.
Therefore, the von Neumann entropy change of the subsystem $\Delta S_1^\mathrm{S}$ is the opposite of the information entropy change generated by the unitary transformation, see Eq.~\eqref{eq:info_entropy}, and the thermodynamic entropy change of the system-bath coupling process reads 
\begin{equation}
     -\Delta S_{1}^\mathrm{info}+\Delta S_{1}^\mathrm{B}\geq 0.
\end{equation}
In the combined process made of the initial unitary transformation on the system and the system-bath coupling, each subsystem's state is reset to the initial state, and the total entropy change is given by the thermodynamic entropy change of the baths $\Delta S_{\alpha}^\mathrm{B}$. In Sec.~\ref{sec:bath_entropy} below, we investigate its properties. 

The relative entropy in Eq.~\eqref{eq:2nd-law-relative-entropy} guides us in finding the conditions under which the equality, i.e. reversibility, is attained: When the system-bath coupling modifies the initial uncorrelated system-bath state weakly, 
\begin{equation}
 \rho_1^\mathrm{S+B}(t) = \rho_1^\mathrm{S}(0)\otimes \rho_1^\mathrm{B}(0) + \delta \rho_1^\text{S+B} \,\Rightarrow\, \left\{\begin{array}{l}
        \rho_1^\text{S/B}(t) = \rho_1^\text{S/B}(0) + \delta \rho_1^\text{S/B}\\
        \delta \rho_1^\text{S/B} := \Tr_\text{B/S}\left\{\delta\rho_1^\text{S+B}\right\}
    \end{array}\right.,
\end{equation}
where $\delta\rho_1^\text{S+B}$ is a traceless matrix, the entropy change in the system and in the bath [Eq.~\eqref{eq:2nd-law-relative-entropy}] sum to zero.
To see this, we write the final state as 
\begin{equation}
    \rho_1^\text{S+B}(t) = \rho_1^\text{S}(t)\otimes \rho_1^\text{B}(t)+ \delta\rho_1^\text{S+B} - \delta \rho_1^\text{S}\otimes \rho_1^\text{B}(0) - \rho_1^\text{S}(0)\otimes \delta \rho_1^\text{B} - \delta \rho_1^\text{S}\otimes \delta\rho_1^\text{B}
\end{equation}
and expand both sides of Eq.~\eqref{eq:2nd-law-relative-entropy} up to lowest non-vanishing order, see Appendix~\ref{app:relative_entropy_expansion}.
Expanding the logarithm and using that $\Tr\left\{\delta \rho_1^\text{S+B}\right\}=0$, we recognize how the first order contribution in $\delta \rho_1^\text{S+B}$ of the relative entropy vanishes, such that  $D\left[\rho_1^\mathrm{S+B}(t)||\rho_1^\mathrm{S}(t)\otimes\rho_1^\mathrm{B}(t)\right] = \smallO[\delta\rho_1^\mathrm{S+B}]$, where $\smallO[\delta\rho_1^\mathrm{S+B}]$ indicates higher than first-order contributions in $\delta \rho_1^\mathrm{S+B}$.
By contrast, the change in entropies entering the right-hand side of Eq.~\eqref{eq:2nd-law-relative-entropy} has a non-vanishing first-order contribution $\Delta S_1^\mathrm{S/B} = -\Tr\left\{\delta\rho_1^\text{S/B}\log \rho_1^\text{S/B}(0)\right\} + \smallO[\delta \rho_1^\text{S+B}]$.
Therefore, when $\delta\rho_1^\text{S+B}$ can be treated as a weak perturbation of the initial state, the changes in entropy satisfy
\begin{equation}
    \Delta S_1^\text{B} = - \Delta S_1^\text{S} = \Delta S_1^\text{info},
\end{equation}
meaning that the information entropy change in the subsystem is fully ``converted" into thermodynamic entropy change of the bath.
This weakly perturbed limit was also investigated using scattering theory in Ref.~\cite{bru2018Mar}, where it is achieved using an adiabatic external drive. While at the lowest (quasistatic) order the information entropy change is found to coincide with the thermodynamic entropy production in the baths, beyond this limit the two are generally different.

\subsection{Entropy production in the bath}\label{sec:bath_entropy}
Previously, we mentioned how the baths we consider are ``much larger" than the subsystems they connect to. Concretely, this is represented by the important assumption that the bath's density operator changes only slightly in the time interval $t$ of the system-bath coupling, such that we can expand it as
\begin{equation}
    \rho_1^\text{B}(t) = \rho_1^\text{B}(0) + \delta\rho_1^\text{B}(t) + \smallO[\delta\rho_1^\text{B}(t)].
\end{equation}
This expansion is justified when the system-bath coupling $\hat{V}_{1}^\text{SB}$ is weak. 
Then, the corresponding expansion of the bath's entropy change reads 
\begin{eqnarray}\label{eq:bath entropy_expand}
\Delta S_{1}^\mathrm{B} & = & -\mathrm{Tr}\left\{\delta\rho_1^\mathrm{B}(t)\log\left[\rho_1^\mathrm{B}(0)\right] \right\} + \smallO[\delta\rho_1^\text{B}(t)]\nonumber\\
& = & -\mathrm{Tr}\left\{\left(\rho_1^\mathrm{B}(t)-\rho_1^\mathrm{B}(0)\right)\log\left[\rho_1^\mathrm{B}(0)\right] \right\} + \smallO[\delta\rho_1^\text{B}(t)],
\end{eqnarray}
which represents the change in expectation value of the operator $\log\left[\rho_1^\mathrm{B}(0)\right]$ due to system-bath coupling.
This operator acquires particular physical significance when the bath is in a thermal state
\begin{equation}
\begin{split}
    \rho_1^\text{B}(0) &= \frac{1}{Z_1} \exp\left[\frac{\hat{H}_1^\text{B} -\mu_1\hat{N}_1^\text{B}}{T_1}\right],\\
    Z_1&= \mathrm{Tr}\left\{\exp\left[\frac{\hat{H}_1^\text{B} -\mu_1\hat{N}_1^\text{B}}{T_1}\right]\right\},
\end{split}
\end{equation}
and allows us to write the entropy production of the bath Eq.~\eqref{eq:bath entropy_expand} in terms of the change in extensive quantities such as the internal energy and the particle number.

In the following we will also consider baths whose occupation across different energies is not described by a unique temperature and chemical potential. We assume that the baths are described by a non-interacting Hamiltonian $\hat{H}_1^{\text{B}} = \sum_k \epsilon_k \hat{N}_1^\text{B}(\epsilon_k)$ where $\epsilon_k$ is the single-particle energy, and $\hat{N}_1^\text{B}(\epsilon_k) = \sum_i \hat{c}_{1,ki}^\dagger \hat{c}_{1,ki} $ is the operator accounting for the particle number at the energy $\epsilon_k$ written in terms of the particle field operators $\hat{c}_{1,ki},\hat{c}_{1,ki}^\dagger$.
We consider the case in which the particles at \textit{each} energy $\epsilon_{k}$ are described by a temperature $T_1(\epsilon_{k})$ and a chemical potential $\mu_1(\epsilon_k)$, such that the bath's state reads
\begin{equation}\label{eq:nonthermal_bath}
\begin{split}
    \rho_1^\mathrm{B}(0)&=\frac{1}{Z_1}\exp\left[\sum_k-\frac{\epsilon_k-\mu_1(\epsilon_k)}{T_1(\epsilon_k)}\hat{N}_1^\text{B}(\epsilon_k)\right],\\
    Z_1 &= \Tr\left\{\exp\left[\sum_k-\frac{\epsilon_k-\mu_1(\epsilon_k)}{T_1(\epsilon_k)}\hat{N}_1^\text{B}(\epsilon_k)\right]\right\}.
\end{split}
\end{equation}
Naturally, when $T_1(\epsilon_k)=T_1$ and $\mu_1(\epsilon_k)=\mu_1$ are energy-independent the state in Eq.~\eqref{eq:nonthermal_bath} reduces to a thermal state.
The average occupation of an energy $\epsilon_k$ is given by the familiar Fermi or Bose distributions (upper or lower sign respectively)
\begin{equation}\label{eq:nonthermal_occupation}
    f_1(\epsilon_k) = \Tr\left\{\hat{c}_{1,ki}^\dagger\hat{c}_{1,ki} \rho_1^\text{B}(0)\right\} = \frac{1}{\exp\left[\frac{\epsilon_k-\mu_1(\epsilon_k)}{T_1(\epsilon_k)}\right] \pm 1}
\end{equation}
depending on the nature of the particles considered, but with energy-dependent temperature and chemical potential. 

For these baths, the thermodynamic entropy production is decomposed according to the energy at which the particle number is changed.
Indeed, combining Eqs.~\eqref{eq:bath entropy_expand} and \eqref{eq:nonthermal_bath} we find
\begin{equation}\label{eq:entropy_particlenumber}
\begin{split}
    \Delta S_{1}^\mathrm{B} &\approx\sum_k \Delta S_1^\mathrm{B}(\epsilon_k)=\sum_k\frac{\epsilon_k-\mu_1(\epsilon_k)}{T_1(\epsilon_k)}\mathrm{Tr}\left\{\left(\rho_1^\mathrm{B}(t)-\rho_1^\mathrm{B}(0)\right)\hat{N}_1(\epsilon_k)\right\}\\
    & =  \sum_k\frac{\epsilon_k-\mu_1(\epsilon_k)}{T_1(\epsilon_k)}\Delta N^\mathrm{B}_1(\epsilon_k)= \sum_k\log\left[\frac{1\mp f_1(\epsilon_k)}{f_1(\epsilon_k)}\right]\Delta N^\mathrm{B}_1(\epsilon_k),
\end{split}
\end{equation}
which depends on the change in particle number at each energy.
To calculate the thermodynamic entropy change, we link the change in the bath observables to the change in the system observables.
In particular, in the following section, we apply this framework to the case of scattering theory, where particles are injected from the baths to the leads (or absorbed by the baths from the leads) without changing their energy.
To capture this property we therefore assume that the system-bath coupling $\hat{V}_1^\text{SB}$ fulfills
\begin{equation}\label{eq:tunnel_strong_condition}
    [\hat{V}_1^\text{SB}, \hat{N}_1^\text{S}(\epsilon_k)+\hat{N}_1^\text{B}(\epsilon_k)]=0,
\end{equation}
which is guaranteed in the weak-coupling regime, where the \textit{energy-resolved} particle conservation is enforced by Fermi's golden rule.
With this, we calculate the particle-number change entering the thermodynamic entropy change of the bath, Eq.~\eqref{eq:entropy_particlenumber}, by evaluating the change in the system's particle number since
\begin{equation}\label{eq:N_strong_condition}
    \Delta N^\mathrm{B}_1(\epsilon_k) = - \Delta N^\mathrm{S}_1(\epsilon_k).
\end{equation}
Note that the \textit{energy-resolved} conditions~\eqref{eq:tunnel_strong_condition} and~\eqref{eq:N_strong_condition} for particle conservation are strictly required only for nonthermal baths, where parameters $\mu(\epsilon_k),T(\epsilon_k)$ can be different at every energy $\epsilon_k$; in the more standard thermal case, overall particle conservation is sufficient.

Up to this point, only the average of the state's observables have been employed.
In the following, we go beyond that by using the two-point measurement scheme to access the underlying probability distribution of the change in the state's observables, such as the change in the particle number.
This allows us to (i) derive scattering theory results for the average and the zero-frequency noise of the currents from a different perspective, and (ii) provide a consistent definition of stochastic entropy production for both information entropy and thermodynamic entropy.
In particular, (i) acts not only as a sanity check, but also as a proof of concept to adapt methods from quantum stochastic thermodynamics to quantum transport.
What is more, (ii) clarifies the difference between the entropy currents defined in scattering theory~\cite{bru2018Mar,deg2020} and gives a formal justification of the definition of entropy current fluctuations, as it has previously been used in Ref.~\cite{Acciai2024Feb}.

\section{Entropy production in a scattering process}\label{sec:Scattering}

In the previous section we have formulated both the information entropy change and the thermodynamic entropy change of a subsystem during the process in terms of the generic input and output states, $\rho_\text{1,in/out}$.
Now, we apply the description above to a scattering problem in which the particles starting in different subsystems evolve unitarily and without interacting with each other.
This allows us to decompose the unitary evolution of the system in terms of the single-particle unitary evolution, and, as we will see in Sec.~\ref{sec:scattering_theory}, leads to the scattering theory formulation of coherent transport.
To highlight this application of the previous section, from now on we associate the subsystems to \textit{leads} of the transport setting shown in Fig.~\ref{fig}(b).

\subsection{Input state of the leads}\label{sec:rho_in}
We consider the $L$ leads of a coherent conductor as subsystems of the general setting discussed before, labeled by greek indices $\alpha=1,\cdots, L$, each with $C_\alpha$ modes, labeled by roman indices $i=1,\cdots, C_\alpha$.
Each lead is described by the Hamiltonian
\begin{equation}
    H_{\alpha} = \sum_{i=1}^{C_\alpha} \epsilon_{\alpha i} c_{\alpha i}^\dagger c_{\alpha i}
\end{equation}
where $c_{\alpha i},c_{\alpha i}^\dagger$ are the particle field operators.
We denote the empty state of the leads as
\begin{equation}
    \ket{\varnothing} \equiv {\ket{0\cdots 0}_1} \cdots \ket{0\cdots 0}_L,
\end{equation}
from which we obtain a state with $k$ particles by applying the creation field operators
\begin{equation}\label{eq:def_vec_alpha i}
    \ket{\boldsymbol{\alpha i}} \equiv c_{\alpha_1i_1}^\dagger \cdots c_{\alpha_k i_k}^\dagger \ket{\varnothing}.
\end{equation}
In the case of fermions, to avoid inconsistencies in the sign of Eq.~\eqref{eq:def_vec_alpha i}, the field operators $c_{\alpha i}$ are ordered lexicographically in the definition of $\ket{\boldsymbol{\alpha i}}$, i.e. $\alpha_xi_x$ appears before $\alpha_y i_y$ if and only if $\alpha_x<\alpha_y$ or $\alpha_x=\alpha_y, i_x<i_y$.
We also remark here that, for fermions, the states $\ket{\boldsymbol{\alpha i}}$ are normalized, i.e. $\braket{\boldsymbol{\alpha i}|\boldsymbol{\alpha i}} = 1$, for any valid state (i.e. states where at most one fermion is in each mode). By contrast, for bosons, these states are not normalized, but instead have norm $\braket{\boldsymbol{\alpha i}|\boldsymbol{\alpha i}} = \prod_{\beta j} n_{\beta j}^{\boldsymbol{\alpha i}}!$, where $n_{\beta j}^{\boldsymbol{\alpha i}}$ is the number of particles in mode $\beta j$ of the state $\ket{\boldsymbol{\alpha i}}$. This normalization factor accounts for the possible permutations of bosons within each mode.
For convenience, we can and will keep the normalization factor $\braket{\boldsymbol{\alpha i}|\boldsymbol{\alpha i}}$ for both fermions and bosons.

With these definitions in place, we construct the initial states of the leads.
Since initially each lead's state is ``prepared" by the corresponding bath independently, we consider a separable input state
\begin{equation}\label{eq:input_state}
    \rho_\text{in} = \bigotimes_\alpha \rho_{\alpha,\text{in}}.
\end{equation}
Each density matrix $\rho_\alpha$---represented by colored circles in Fig.~\ref{fig}---is taken to be a mixed state of the form 
\begin{equation}\label{eq:marginal_input_state}
    \rho_{\alpha,\text{in}} = \sum_{n_1,\cdots, n_{C_\alpha}} p_{\alpha 1}(n_1)\cdots p_{\alpha C_\alpha}(n_{C_\alpha}) \frac{\proj{n_1\cdots n_{C_\alpha}}_\alpha}{n_1!\cdots n_{C_\alpha}!},
\end{equation}
where $p_{\alpha i}(n_i)$ is the probability of having $n_i$ particles in channel $i$ of lead $\alpha$. These probabilities are determined by the bath's statistics.
Concretely, when the baths are described by the states Eq.~\eqref{eq:nonthermal_bath}, we write the probabilities $p_{\alpha i}(n_i)$ in terms of the average occupation of Eq.~\eqref{eq:nonthermal_occupation} as 
\begin{subequations}\label{eq:fermi-bose-statistics}
 \begin{alignat}{2}
     p_{\alpha i}(1)&=1-p_{\alpha i}(0) = f_\alpha(\epsilon_{\alpha i})\qquad &&\text{for fermions,} \\
     p_{\alpha i}(n) &= \frac{1}{f_\alpha(\epsilon_{\alpha i})+1}\left[\frac{f_\alpha (\epsilon_{\alpha i})}{f_\alpha(\epsilon_{\alpha i})+1}\right]^{n} \qquad &&\text{for bosons}.
 \end{alignat}   
\end{subequations}
Note that, for each lead to equilibrate with the corresponding bath and to inherit its occupation probabilities, the system-bath coupling $\hat{V}_\alpha^\text{SB}$ needs to be weak.

\subsection{Output state of the leads}\label{sec:rho_out}
The input state $\rho_\text{in}$ undergoes a unitary evolution mapping it into the output state $\rho_\text{out}$.
In the scattering process we are now considering, this unitary evolution is fully determined by the single-particle transformation
\begin{equation}\label{eq:scattering-matrix-relation}
    {c}^\dagger_{\alpha i}\to \sum_{\beta j} s_{\beta j, \alpha i} {c}^\dagger_{\beta j},
\end{equation}
where $s$ is a $(\sum_\alpha C_\alpha)\times(\sum_\alpha C_\alpha)$ unitary matrix and is referred to as \textit{scattering matrix}.
Simply put, the element $s_{\beta j, \alpha i}$ is the amplitude with which an input particle in $\alpha i$ ends in $\beta j$ in the output.
By applying the single-particle transformation Eq.~\eqref{eq:scattering-matrix-relation} to each field operator in a given state, see Eq.~\eqref{eq:def_vec_alpha i}, we find the corresponding output state, namely
\begin{equation}
\label{eq:initial-final-state}
\ket{\boldsymbol{\alpha i}}\to 
\ \prod_{x=1}^{k} \left[ \sum_{\beta j} s_{\beta j,\alpha_xi_x} {c}^\dagger_{\beta j}\right] \ket{\varnothing} = \sum_{\boldsymbol{\beta j}} \text{det}_\mp[\mathfrak{S}^{\boldsymbol{\beta j}}_{\boldsymbol{\alpha i}}] \frac{\ket{\boldsymbol{\beta j}}}{\braket{\boldsymbol{\beta j}|\boldsymbol{\beta j}}}.
\end{equation}
Here, the summation is done over the different states $\sum_{\boldsymbol{\beta j}} \bullet=\sum_{\beta_1j_1\leq \cdots\leq\beta_kj_k} \bullet $, where $\det_{-}[\bullet]$ is the determinant emerging from the exchange statistics of fermions and $\det_{+}[\bullet]$ is the permanent emerging from the exchange statistics of bosons. The determinant/permanent is evaluated on the matrix $\mathfrak{S}^{\boldsymbol{\beta j}}_{\boldsymbol{\alpha i}}$ with elements
\begin{equation}
{(\mathfrak{S}^{\boldsymbol{\beta j}}_{\boldsymbol{\alpha i}})}_{xy} := s_{\beta_xj_x, \alpha_y i_y}
\end{equation}
encoding all the possible ways to map state $\ket{\boldsymbol{\alpha i}}$ into state $\ket{\boldsymbol{\beta j}}$.
This transformation generates correlations between the different leads, as depicted in Fig.~\ref{fig} by the purple lines, and can be used to produce entangled electronic states~\cite{Beenakker2006,Klich2009Mar,Hofer2016Dec}.

Using Eq.~\eqref{eq:initial-final-state}, we can transform the density matrix associated to the state $\ket{\boldsymbol{\alpha i}}$ into the corresponding output state $\rho_{\text{out}|\boldsymbol{\alpha i}}$
\begin{equation}
\frac{\proj{\boldsymbol{\alpha i}}}{\braket{\boldsymbol{\alpha i}|\boldsymbol{\alpha i}}} \to \rho_{\text{out}|\boldsymbol{\alpha i}}. 
\end{equation}
This is particularly useful since the input state $\rho_\text{in}$ discussed in the previous section is a convex combination of projectors $\frac{\proj{\boldsymbol{\alpha i}}}{\braket{\boldsymbol{\alpha i}|\boldsymbol{\alpha i}}}$.
Indeed, the output state $\rho_\text{out}$ is given by averaging the output states conditioned on a specific input state $\boldsymbol{\alpha i}$, namely
\begin{equation}\label{eq:output_state}
    \rho_\text{out} = \sum_{\boldsymbol{\alpha i}}\rho_{\text{out}|\boldsymbol{\alpha i}} p(\boldsymbol{\alpha i}) = \sum_{\boldsymbol{\alpha i}}\rho_{\text{out}|\boldsymbol{\alpha i}} \Tr\left\{\frac{\proj{\boldsymbol{\alpha i}}}{\braket{\boldsymbol{\alpha i}|\boldsymbol{\alpha i}}} \rho_\text{in}\right\}
\end{equation}
where the probability of each input state fulfills $p(\boldsymbol{\alpha i}) = \prod_{\beta j}p_{\beta j}(n^{\boldsymbol{\alpha i}}_{\beta j})$, following from Eq.~\eqref{eq:marginal_input_state}.

The structure of Eq.~\eqref{eq:output_state} as conditional state times probability of input state $\boldsymbol{\alpha i}$ suggests to introduce measurements to sample the statistics of the process. We do so in Sec.~\ref{subsec:2PM}, where we use the two-point measurement scheme to define the stochastic entropy production at different steps of the process.

\subsection{Marginal states and average entropy changes}

In Sec.~\ref{sec:info-thermo-entropy} we have seen how the marginal states on a given lead are required to calculate both the information [Eq.~\eqref{eq:info_entropy}] and thermodynamic [Eq.~\eqref{eq:bath entropy_expand}] entropy changes of the process.
Once again, we focus without loss of generality on lead 1.

The marginal input state of lead 1 is given in Eq.~\eqref{eq:marginal_input_state} (setting $\alpha=1$) since we have chosen the input state to be separable over the different leads, see Eq.~\eqref{eq:input_state}.
To obtain the marginal output state, we need to trace out all other leads, here indicated as $\bar{1}$, from the output state in Eq.~\eqref{eq:output_state}. The decomposition of the input state into the projectors $\proj{\boldsymbol{\alpha i}}$ allows us to write
\begin{equation}\label{eq:marginal_output_state_decomposition}
    \rho_{1,\text{out}} =  \Tr_{\bar{1}}\left\{\rho_\text{out}\right\} = \sum_{\boldsymbol{\alpha i}} \Tr_{\bar{1}}\left\{\rho_{\text{out}|\boldsymbol{\alpha i}}\right\}p(\boldsymbol{\alpha i}) = \sum_{\boldsymbol{\alpha i}} \rho_{\text{1}|\boldsymbol{\alpha i}}p(\boldsymbol{\alpha i})
\end{equation}
where we called $\rho_{\text{1}|\boldsymbol{\alpha i}}:=\Tr_{\bar{1}}\left\{\rho_{\text{out}|\boldsymbol{\alpha i}}\right\}$ the marginal output state of lead 1 \textit{conditioned} on a specific initial state $\ket{\boldsymbol{\alpha i}}$. This marginal state has the block structure
\begin{equation}
    \rho_{1|\boldsymbol{\alpha i}} = \begin{pmatrix} \rho_{1(0)|\boldsymbol{\alpha i}} & 0 &0 & \cdots\\
0&\rho_{1(1)|\boldsymbol{\alpha i}}& 0 & \cdots\\
0 & 0 &\rho_{1(2)|\boldsymbol{\alpha i}}&\cdots\\
\vdots&\vdots &\vdots&\ddots
\end{pmatrix}    
\end{equation}
with block $\rho_{1(n)|\boldsymbol{\alpha i}}$ containing $n$ particles in lead 1. Clearly, since this state is conditioned on the initial state and the number of particles is conserved in the process, the maximum $n$ for which $\rho_{1(n)|\boldsymbol{\alpha i}} \neq 0$ corresponds to the number $k$ of particles in the initial state $\ket{\boldsymbol{\alpha i}}$.
Concretely, the block with $n$ particles is given by
\begin{equation}
    \rho_{1(n)|\boldsymbol{\alpha i}} =\frac{1}{\braket{\boldsymbol{\alpha i}|\boldsymbol{\alpha i}}} \sum_{\boldsymbol{1x},\boldsymbol{1y}} \sum_{1C_1<\boldsymbol{\zeta q}} \frac{ \mathrm{det}_\mp[\mathfrak{S}_{\boldsymbol{\alpha i}}^{\boldsymbol{1x\zeta q}}]\mathrm{det}_\mp[\mathfrak{S}_{\boldsymbol{\alpha i}}^{\boldsymbol{1y\zeta q}}]^*}{\braket{\boldsymbol{\zeta q}|\boldsymbol{\zeta q}}} \frac{\ketbra{\boldsymbol{1x}}{\boldsymbol{1y}}}{\braket{\boldsymbol{1x}|\boldsymbol{1x}}\braket{\boldsymbol{1y}|\boldsymbol{1y}}},
\end{equation}
where $\boldsymbol{1x\zeta q}=(1x_1,1{x_2},\cdots, 1{x_n},\zeta_1q_1,\cdots, \zeta_{k-n}q_{k-n})$ is ordered and contains $n$ particles in the first lead, while the remaining $k-n$ are not in the first lead, as specified by the sum $\sum_{1C_1<\boldsymbol{\zeta q}}$.

With the marginal input [Eq.~\eqref{eq:marginal_input_state}] and the marginal output state [Eq.~\eqref{eq:marginal_output_state_decomposition}] we calculate the von Neumann entropy entering the information-entropy change of Eq.~\eqref{eq:info_entropy}, namely
\begin{subequations}
\begin{align}
    S[\rho_\text{1,in}] &= -\sum_{\boldsymbol{1z}} p(\boldsymbol{1z})\log p(\boldsymbol{1z}), \\
    S[\rho_\text{1,out}]&=-\Tr\left\{\rho_\text{1,out}\log \rho_\text{1,out}\right\} \geq \sum_{\boldsymbol{\alpha i}} p(\boldsymbol{\alpha i})S[\rho_{1|\boldsymbol{\alpha i}}].
\end{align}
\end{subequations}
Note that the information entropy change $\Delta S_1^\text{info} = S[\rho_\text{1,out}] - S[\rho_\text{1,in}]$ is not linear in the transmission function of the conductor.
By contrast, the thermodynamic entropy change of the bath $\Delta S_1^\text{B}$, discussed in Sec.~\ref{sec:bath_entropy}, depends on the change of physical observables such as energy and particle number, and is therefore linear in the conductor's transmission function.
The difference between these \textit{average} entropy changes has also been pointed out in Refs.~\cite{bru2018Mar, esp2015, ses2021feb, zhou2024feb}. Here, we go beyond this picture to access the \textit{fluctuations} of such quantities. To do so, we introduce in Sec.~\ref{subsec:2PM} a two-point measurement scheme on the lead occupations before and after the scattering process.

\subsection{Two-point measurement scheme and stochastic entropy production}\label{subsec:2PM}
The decomposition of the marginal output state $\rho_{\text{out}}$ in Eq.~\eqref{eq:output_state} is an ideal starting point to use a two-point measurement scheme to describe the fluctuations in the scattering process.

The measurements are described by the complete set of projectors $\left\{\hat{\Pi}_{\boldsymbol{\alpha i}}:=\frac{\proj{\boldsymbol{\alpha i}}}{\braket{\boldsymbol{\alpha i}|\boldsymbol{\alpha i}}}\right\}_{\boldsymbol{\alpha i}}$ measuring the number of particles in each mode of all leads. The input state is already diagonal in this basis, so the first measurement does not perturb the state, and yields outcome $\boldsymbol{\alpha i}$ with probability $P(\boldsymbol{\alpha i}) = p(\boldsymbol{\alpha i})$ [see Eq.~\eqref{eq:output_state}]. The second measurement has outcome $\boldsymbol{\zeta z}$ with probability
\begin{equation}\label{eq:conditional-2PM-prob}
    P(\boldsymbol{\zeta z}|\boldsymbol{\alpha i}) = \Tr\left\{\frac{\proj{\boldsymbol{\zeta z}}}{\braket{\boldsymbol{\zeta z}|\boldsymbol{\zeta z}}}\rho_{\text{out}|\boldsymbol{\alpha i}}\right\} = \frac{\left|\text{det}_\mp \left[\mathfrak{S}_{\boldsymbol{\alpha i}}^{\boldsymbol{\zeta z}}\right]\right|^2}{\braket{\boldsymbol{\alpha i}|\boldsymbol{\alpha i}}\braket{\boldsymbol{\zeta z}|\boldsymbol{\zeta z}}}
\end{equation}
conditioned on the outcome $\boldsymbol{\alpha i}$ of the first measurement.
Note that, unlike the first measurement, this second measurement destroys the coherences created by the scattering process, as shown in Fig.~\ref{fig}, and therefore affects the observed entropy.

From the conditional probability Eq.~\eqref{eq:conditional-2PM-prob}, we define the stochastic entropy production that emerges from the two-point measurement scheme (TPM) as~\cite{Esposito2009Dec, Landi2021Sep, Strasberg2022Jan}
\begin{equation}\label{eq:stoch_TPM_entropy_def}
    \sigma^\text{TPM}(\boldsymbol{\zeta z};\boldsymbol{\alpha i}) := -\log P(\boldsymbol{\zeta z}|\boldsymbol{\alpha i}) + \log P(\boldsymbol{\alpha i}).
\end{equation}
Its expectation value then reads
\begin{equation}\label{eq:stoch_TPM_entropy}
\begin{split}
    \mathbb{E}[\sigma^\text{TPM}(\boldsymbol{\zeta z};\boldsymbol{\alpha i})] &= \Shannon[\boldsymbol{\zeta z}|\boldsymbol{\alpha i}]-\Shannon[\boldsymbol{\alpha i}] \\
    &= \Shannon[\boldsymbol{\zeta z}] - \Shannon[\boldsymbol{\alpha i}] - I[\boldsymbol{\zeta z}:\boldsymbol{\alpha i}]
\end{split}
\end{equation}
where $\Shannon[X]:=-\sum_X P(X)\log P(X)$ is the Shannon entropy of the probability distribution $P(X)$, $\Shannon[X|Y]:=-\sum_{X,Y}P(X,Y)\log [P(X|Y)]$ is the conditional entropy, and $I[\boldsymbol{\zeta z}:\boldsymbol{\alpha i}]:= \Shannon[\boldsymbol{\zeta z}] - \Shannon[\boldsymbol{\zeta z}|\boldsymbol{\alpha i}]$ is the mutual information between the first and second measurement outcomes.
Noticing that the Shannon entropies in Eq.~\eqref{eq:stoch_TPM_entropy} coincide with the Von Neumann entropies of the measured input and output states, and using that the first measurement does not perturb the input state $\mathcal{M}^\text{TPM}\rho_\text{in} = \sum_{\boldsymbol{\alpha i}}\hat{\Pi}_{\boldsymbol{\alpha i}}\rho_\text{in}\hat{\Pi}_{\boldsymbol{\alpha i}} = \rho_\text{in}$, the average TPM entropy production in Eq.~\eqref{eq:stoch_TPM_entropy} becomes
\begin{equation}\label{eq:stoch_TPM_entropy_avg}
\begin{split}
    \mathbb{E}[\sigma^\text{TPM}(\boldsymbol{\zeta z};\boldsymbol{\alpha i})] &= S[\mathcal{M}^\text{TPM}\rho_\text{out}] - S[\rho_\text{in}]- I[\boldsymbol{\zeta z}:\boldsymbol{\alpha i}]\\
    &\geq - I[\boldsymbol{\zeta z}:\boldsymbol{\alpha i}]
\end{split}
\end{equation}
where the inequality is obtained by using that $S[\mathcal{M}^\text{TPM}\rho]\geq S[\rho]$ since $\mathcal{M}^\text{TPM}$ is a unital map.
In contrast with the information entropy of Eq.~\eqref{eq:info_entropy}, in the TPM entropy production the coherences of the output state are destroyed, but classical correlations between different leads are still present.
This is a consequence of the measurements acting on \textit{all} leads.
To account for \textit{local} measurements, we restrict the two-point measurement scheme to measurements performed to one lead only. Then, instead of the joint probability distribution $P(\boldsymbol{\zeta z};\boldsymbol{\alpha i})$, one would sample the coarse-grained probability distribution $P(\boldsymbol{1y};\boldsymbol{1x}) = \sum_{1C_1<\boldsymbol{\zeta z}}\sum_{1C_1<\boldsymbol{\alpha i}}P(\boldsymbol{1y\zeta z};\boldsymbol{1x\alpha i})$.
Then, the TPM entropy production on lead 1 reads
\begin{equation}\label{eq:stoch_TPM1_entropy_def}
     \sigma^\text{TPM}_1(\boldsymbol{y};\boldsymbol{x}) := -\log P(\boldsymbol{1y}|\boldsymbol{1x}) + \log P(\boldsymbol{1x}),
\end{equation}
mirroring Eq.~\eqref{eq:stoch_TPM_entropy_def}. To evaluate its expectation value, we once again identify the Shannon entropies with the corresponding von Neumann entropies of the measured input and output \textit{reduced} states on lead 1.
Then, since the input state is not perturbed by the measurement, $\mathcal{M}_1^\text{TPM}\rho_{1, \text{in}} = \sum_{\boldsymbol{1x}} \hat{\Pi}_{\boldsymbol{1x}}\rho_{1,\text{in}}\hat{\Pi}_{\boldsymbol{1x}} = \rho_{1, \text{in}}$, we have
\begin{equation}\label{eq:stoch_TPM1_entropy_avg}
    \begin{split}
    \mathbb{E}[\sigma^\text{TPM}_1(\boldsymbol{y};\boldsymbol{x})] &=S[\mathcal{M}^\text{TPM}_1\rho_\text{1,out}] - S[\rho_\text{1,in}]- I[\boldsymbol{1y}:\boldsymbol{1x}]\\
    &\geq \Delta S_1^\text{info}- I[\boldsymbol{1y}:\boldsymbol{1x}],
\end{split}
\end{equation}
just as in Eq.~\eqref{eq:stoch_TPM_entropy_avg} before.
Here, how the measurement affects the information gained on the system compared to the entropy change Eq.~\eqref{eq:info_entropy} is more apparent: On the one hand, the destruction of the coherences in the reduced output state increases the entropy production; on the other hand, keeping track of the initial measurement outcome reduces the uncertainty of the process, and is accounted for by the mutual information $I[\boldsymbol{1y}:\boldsymbol{1x}]$.

The difference between the entropy changes of Eqs.~(\ref{eq:von_Neumann_entropy_conserved}, \ref{eq:info_entropy}, \ref{eq:stoch_TPM_entropy_avg}, \ref{eq:stoch_TPM1_entropy_avg}) showcases how the information entropy depends on the level of description of the leads and on the measurements adopted to sample the system.
This is not the case for the thermodynamic entropy instead because the measurement does not change the average number of transferred particles.
Having access to the single realizations of the scattering process, we also know the net amount of particles flowing into each bath with each event.
Since this net amount is not a fixed value but a stochastic one, the entropy produced in the baths due to single events is also a stochastic variable. Following Eq.~\eqref{eq:entropy_particlenumber}, the stochastic entropy change of bath 1 is
\begin{equation}\label{eq:stoch_thermo_entropy}
    \sigma^\text{B}_1(\boldsymbol{\zeta z}|\boldsymbol{\alpha i}) = \sum_{x=1}^{C_1}\log\left[\frac{1\mp f_1(\epsilon_{1x})}{f_1(\epsilon_{1x})}\right] (n_{1x}^{\boldsymbol{\zeta z}} - n_{1x}^{\boldsymbol{\alpha i}}),
\end{equation}
such that its average coincides with the thermodynamic entropy production in the bath,
\begin{equation}
    \Delta S_1^\text{B} = \mathbb{E}[\sigma^\text{B}_1(\boldsymbol{\zeta z}|\boldsymbol{\alpha i})].
\end{equation}
Importantly, the stochastic entropy production Eq.~\eqref{eq:stoch_thermo_entropy} allows us to consider also higher cumulants of the entropy production.
Naturally, this protocol is not limited to entropy, but also applies to other stochastic quantities of interest, such as the particle number of the energy that is transferred in a bath.

\section{Currents and noise in scattering theory}\label{sec:scattering_theory}

In the previous section we described a single scattering event, and used the two-point measurement scheme to define the change in observables between the input and the output states of the scattering process.
We now take this as the starting point to obtain the same formulations for the currents in scattering theory by ``counting" how often the single scattering events take place.
This approach not only allows us to obtain the same statistics for the particle current, but also to provide well-defined entropy currents and entropy current noise based on the stochastic entropy productions introduced in Sec.~\ref{subsec:2PM}.

Calling $v_{1}(\epsilon_{1 x})$ the particle flux in lead $1$ and at energy $\epsilon_{1 x}$, we write the stochastic current contribution of mode $1x$ as
\begin{equation}
\mathcal{J}_{1 x}^{(O)} = v_1(\epsilon_{1x})o_{1x}\delta n_{1x},
\end{equation}
where $o_{1x}$ is the generalized charge carried in mode $1x$, while $\delta n_{1x}$ is the change in the number of particles in mode $1x$ in the scattering process, i.e.
\begin{equation}
    \delta n_{1x} = n_{1x}^{\boldsymbol{\zeta z}} - n_{1x}^{\boldsymbol{\alpha i}}.
\end{equation}
To explicitly include in the description the possibility of having multiple channels at the same energy in the leads, we now specify the index to be $x=(l,q)$. Here, the first index determines the energy of the considered channels, whereas the second index distinguishes between degenerate channels at the same energy.
Concretely, this means 
\begin{equation}\label{eq:index_replacement}
\epsilon_{1x} = \epsilon_{1 l},\quad \delta n_{1x} = \delta n_{1 q}(\epsilon_{1 l}).
\end{equation}
Then, the stochastic current generated by the scattering event reads
\begin{equation}
\begin{split}    
    \mathcal{J}_{1}^{(O)} &= \sum_x\mathcal{J}^{(O)}_{1x} = \int dE \sum_{lq} v_1(E)o_{1q}(E)\delta n_{1 q}(E) \delta (E - \epsilon_{1 l}) \\
    &=\int dE \sum_{q} v_1(E)g_1(E)o_{1q}(E)\delta n_{1 q}(E),
\end{split}
\end{equation}
where we introduced the density of states $g_1(E):=\sum_{l}\delta(E-\epsilon_{1 l})$.
In one-dimensional leads the density of states is $g_1(E) = [hv_1(E)]^{-1}$~\cite{Blanter2000Sep}.
Furthermore, we approximate the flux as energy-independent, $v_1(E)=v_1$.
Then, the average current into lead $1$ reads
\begin{equation}\label{eq:average-current}
    J_1^{(O)} = \int \frac{dE}{h} \sum_{q} o_{1q}(E)\mathbb{E}[\delta n_{1 q}(E)] = \int \frac{dE}{h}\sum_{\beta qq'l} o_{1q}(E) |s_{1q,\beta q'}(E,E_l)|^2 [f_\beta(E_l) - f_1(E)],
\end{equation}
which is the Landauer-B\"uttiker formula~\cite{Blanter2000Sep, Moskalets2011Sep} for the transport of the quantity $O$, see Appendix~\ref{app:probability-distribution} for the technical details.
Here, the scattering matrix $s_{1q, \beta q'}(E,E_{l})$ also includes transitions from energy $E_l$ to energy $E$, which is the case in driven conductors, where particles exchange energy quanta with the drive.
In this case, the current in Eq.~\eqref{eq:average-current} corresponds to the time-averaged current, as detailed in Appendix~\ref{app:scaled-cumulants}.

Additionally, the statistical independence between different scattering events allows us to write the zero-frequency noise of the particle current in lead 1 in terms of the variance of the transferred quantity $O$
\begin{equation}
    \S_{11}^{(O)} = v_1 \Var{\sum_x o_{1x}\delta n_{1x}}.
\end{equation}
The full expression for the variance is given in Eq.~\eqref{app:eq:Ovar}. Here, we discuss the simpler case in which the scattering matrix does not change the particle energy, and degenerate channels have the same occupation and carry the same quantity $o_{1q}=o_1$. This happens in stationary conductors, where the noise then reads
\begin{equation}
\begin{split}
    \S_{11}^{(O)} = \int \frac{dE}{h}&o_1^2\left(\sum_{\beta\neq 1}\Tr\left\{\boldsymbol{t}_{1\beta}^\dagger \boldsymbol{t}_{1\beta}\right\}\left(f_1[1\mp {f}_\beta] + [1\mp f_1]f_\beta\right)+ \right.\\
    &\left.\qquad \mp \sum_{\beta,\gamma}[f_\beta -f_1][f_\gamma -f_1] \Tr\left\{\boldsymbol{t}_{1\beta} \boldsymbol{t}_{1\beta}^\dagger\boldsymbol{t}_{1\gamma} \boldsymbol{t}_{1\gamma}^\dagger\right\}\right).
\end{split}
\end{equation}
Here, we dropped the energy dependence in the integrand for conciseness, and introduced the transmission matrix $(\boldsymbol{t}_{1\beta})_{qq'} = s_{1q,\beta q'}(E,E)$ mapping the channels in lead $\beta$ into those in lead 1 at the same energy $E$ (so the traces in the previous equation are performed over the channel degree of freedom).
We recover the particle current noise~\cite{Blanter2000Sep} by setting $o_1=1$, meaning that the considered quantity is just the change in particle number due to the scattering process. Equally, energy current and its noise are obtained by replacing $o_1=E$. Importantly, this strategy also allows us to obtain currents and their statistics that are typically not available from scattering theory. 
In particular, when considering the stochastic thermodynamic entropy production Eq.~\eqref{eq:stoch_thermo_entropy},  we set $o_1 = \log\left[\frac{1\mp f_1(E)}{f_1(E)}\right]$ and obtain the entropy current noise used in Ref.~\cite{Acciai2024Feb}.
While the expression for the entropy current noise had previously been intuitively argued via the information carried by particles, the approach presented here provides a rigorous derivation. Importantly, a similar procedure can be envisioned for other observables that are accessible via the system's density operator and brought in connection with particle transfer to the baths.

\section{Conclusions}\label{sec:conclusions}

We studied entropy production and its fluctuations in coherent quantum transport by combining elements of the repeated interaction framework with the two-point measurement scheme in the context of scattering theory.
We distinguish between \textit{information-entropy} change, which arises in the unitary transformation of the system---the scattering process---when only partial information is accessed, and the \textit{thermodynamic entropy change}, which arises in the lead-bath coupling that resets the lead state.
The former was also identified in Refs.~\cite{bru2018Mar,zhou2024feb}, and is generally nonlinear in the transmission probability, whereas the latter fulfills Clausius relation for thermal baths and is linear in the transmission probability~\cite{Esposito2010Jan}.
By employing the two-point measurement scheme, we provided a consistent stochastic description of entropy production at the level of single events, which also supplies a rigorous derivation of the entropy fluctuations as they were previously used in Ref.~\cite{Acciai2024Feb}. Importantly, the developed process can be extended to thermodynamically relevant observables that are obtained from the density operator of the system and features which can be detected in transport currents into the baths.

The framework developed here opens the way to adapting results from quantum stochastic thermodynamics to coherent transport, ranging from studying how the thermodynamics of quantum scattering~\cite{Jacob2021Apr, Jacob2022Jun, Jacob2023Nov} affects transport, to investigating the limits on precision set by thermodynamic uncertainty relations.

The opportunity offered by this framework can be further broadened by relaxing some of the assumptions made throughout the paper. 
First, note that the measurement protocol we used reproduces the zero-frequency noise. However, at finite noise frequencies, the picture is more complicated, and the fluctuations observed depend on the measurement protocol adopted~\cite{Lesovik1997Feb, Glattli2009Jun}. Implementing different measurement protocols to access different combinations of noise, and clarifying their connection to the information gained by the measurement will shine further light on the thermodynamics of out-of-equilibrium fluctuations.
Second, the measurements we considered here did not affect significantly the dynamics: The initial state is unperturbed by the measurement~\cite{Shelankov2003Aug, Esposito2009Dec}, and the measurement does not change the average energy of the state.
Therefore, considering different measurement protocols and including the possibility of feedback, especially on the unitary transformation applied to the system, will allow to consider quantum machines such as measurement engines~\cite{Elouard2017Mar, Elouard2018Jun} using coherent transport systems as platforms.
Third, here we limited the discussion to unitary transformations that decompose into single-particle processes. Extending the framework to more general unitary transformations, which also include particle-particle interactions, as well as to more structured or correlated nonthermal baths, would broaden the applicability of the approach to different transport regimes.

\section*{Acknowledgements}

L.T. and J.S. thank J. Goold and D. Palmqvist for helpful discussions.


\paragraph{Funding information}
 We acknowledge financial support from the Knut and Alice Wallenberg foundation through the fellowship program (L.T.,M.A.,J.S.), from the European Research Council (ERC) under the European Union’s Horizon Europe research and innovation program (101088169/NanoRecycle) (H.K.,J.S.), and the European Union’s Horizon Europe research and
innovation programme under the Marie Sklodowska-Curie grant agreement No. 101205255 FLUTE (M.A.).

\appendix

\section{Expansion of the relative entropy}\label{app:relative_entropy_expansion}

In this Appendix we provide the detailed expansions of relative entropy and entropy changes discussed in Sec.~\ref{sec:thermodynamic-entropy}.

Using that the logarithm of a matrix $A$ can be cast into the series
\begin{align}
    \log{A}=\sum_{k=1}^\infty  \frac{(-1)^{(k+1)}}{k}(A-\mathbbm{1})^k,
\end{align}
we have 
\begin{equation}\label{app:eq:log_expansion}
    \begin{split}
    \log{({\rho}+\delta {\rho})}&=\sum_{k=1}^\infty  \frac{(-1)^{(k+1)}}{k}(\rho -\mathbbm{1}+\delta \rho )^k \\
     &=\sum_{k=1}^\infty  \frac{(-1)^{(k+1)}}{k}[({\rho}-\mathbbm{1})^k+\sum_{l=0}^{k-1}(\rho -\mathbbm{1})^l\delta {\rho}(\rho -\mathbbm{1})^{k-1-l}]+\smallO(\delta \rho ) \\
     &=\log{\rho }+\sum_{k=1}^\infty  \frac{(-1)^{(k+1)}}{k}\sum_{l=0}^{k-1}(\rho -\mathbbm{1})^l\delta {\rho}(\rho -\mathbbm{1})^{k-1-l}+\smallO(\delta \rho ).
    \end{split}
\end{equation}
We therefore expand the relative entropy $D[\rho||\rho+\delta \rho]$ as
\begin{equation}\label{app:eq:relative_entropy_exp}
\begin{split}
    D[\rho||\rho+\delta\rho]&=\Tr\{\rho[\log{\rho }-\log{(\rho +\delta\rho )}]\}\\
    &=-\Tr\bigg\{ \rho  \sum_{k=1}^\infty  \frac{(-1)^{(k+1)}}{k}\sum_{l=0}^{k-1}(\rho -\mathbbm{1})^l\delta {\rho}(\rho -\mathbbm{1})^{k-1-l}\bigg\}+\smallO(\delta \rho) \\
     &=   -\Tr\bigg\{ \rho  \sum_{k=1}^\infty  \frac{(-1)^{(k+1)}}{k}k\delta \rho (\rho -\mathbbm{1})^{k-1}\bigg\}+\smallO(\delta \rho) \\
     &=   -\Tr\{\delta \rho  \}+\smallO(\delta \rho) =\smallO(\delta \rho),
\end{split}
\end{equation}
where we have used the cyclic property of the trace, recognized the series of 
\begin{align}
    \frac{1}{\rho }=\frac{1}{\mathbbm{1}+(\rho -\mathbbm{1})}=\sum_{k=0}^\infty (-1)^k(\rho -\mathbbm{1})^k,
\end{align}
and used that $\Tr\{\delta \rho \}=0$.

In Sec.~\ref{sec:thermodynamic-entropy}, we see that the relative entropy $D[\rho_1^\text{S+B}(t)||\rho_1^\text{S}\otimes \rho_1^\text{B}]$ vanishes at first order in $\delta\rho_1^\text{S+B}$ by replacing
\begin{equation}
    \begin{split}
        \rho &\to \rho_1^\text{S+B}(t),\\
        \delta \rho&\to \delta\rho_1^\text{S}\otimes\rho_1^\text{B}(0) +\rho_1^\text{S}(0)\otimes\delta \rho_1^\text{B} + \delta\rho_1^\text{S}\otimes \delta \rho_1^\text{B} - \delta \rho_1^\text{S+B},
    \end{split}
\end{equation}
in Eq.~\eqref{app:eq:relative_entropy_exp}.

Similarly, we use Eq.~\eqref{app:eq:log_expansion} to expand the entropy change as
\begin{equation}
\begin{split}
    S[\rho+\delta\rho] - S[\rho] = -\Tr\{\delta\rho\log\rho\} + \smallO(\delta \rho),
\end{split}
\end{equation}
which corresponds to the first order contribution to $\Delta S_1^\text{S/B}$ when one chooses $\rho\to \rho_1^\text{S/B}(0)$ and $\delta \rho\to \delta \rho_1^\text{S/B}$.

Therefore, while the relative entropy vanishes at first order in the perturbation, the individual entropy changes of Eq.~\eqref{eq:2nd-law-relative-entropy} do not.
\section{Probability distribution of the two-point measurement}\label{app:probability-distribution}

We now use the conditional probability of Eq.~\eqref{eq:conditional-2PM-prob} to calculate the expectation value and variance of the number of transferred particles, and of stochastic observables generated by the transfer of particles.
These derivations allow us to connect the statistics of a single scattering event as described in Sec.~\ref{sec:Scattering} with the current statistics of Sec.~\ref{sec:scattering_theory}.
Here, we exploit the structure of the conditional probability and do some combinatorics. Alternatively, in Appendix~\ref{app:FCS}, we show that the moment generating function of the two-point measurement scheme leads to the well-known full counting statistics for scattering theory.

In the following, to count over states, it is convenient to consider the probability of an unordered $k$-tuple $(\zeta_1z_1,\cdots,\zeta_kz_k)$. Thus, noticing that the sum over states (or equivalently ordered $k$-tuples) can be unraveled as
\begin{equation}
    \sum_{\boldsymbol{\zeta z}} \bullet = \sum_{\zeta_1 z_1}\cdots \sum_{\zeta_k z_k} \frac{\braket{\boldsymbol{\zeta z}|\boldsymbol{\zeta z}}}{k!} \bullet,
\end{equation}
the unordered probability reads
\begin{equation}\label{app:eq:unordered-conditional-probability}
    P(\zeta_1 z_1,\cdots \zeta_kz_k|\boldsymbol{\alpha i}) = \frac{\left|\text{det}_\mp \left[\mathfrak{S}^{\boldsymbol{\zeta z}}_{\boldsymbol{\alpha i}}\right]\right|^2}{k!\braket{\boldsymbol{\alpha i}|\boldsymbol{\alpha i}}}.
\end{equation}
Exploiting the determinant/permanent structure of the conditional probability, we notice that coarse-graining over the last particle state leads to
\begin{equation}\label{app:eq:probability-property}
\begin{split}
    \sum_{\zeta_k z_k} &P(\zeta_1z_1,\cdots, \zeta_kz_k |\boldsymbol{\alpha i})=\\
    &=\frac{1}{k!\braket{\boldsymbol{\alpha i}|\boldsymbol{\alpha i}}} \sum_{\ell_1,\ell_2=1}^k (\mp1)^{\ell_1+\ell_2} \delta_{\alpha_{\ell_1}i_{\ell_1},\alpha_{\ell_2}i_{\ell_2}} \text{det}_\mp\left[\mathfrak{S}_{\boldsymbol{\alpha i}\setminus\{\ell_1\}}^{\boldsymbol{\zeta z}\setminus\{k\}}\right]\text{det}_\mp\left[\mathfrak{S}_{\boldsymbol{\alpha i}\setminus\{\ell_2\}}^{\boldsymbol{\zeta z}\setminus\{k\}}\right]^*\\
    &=\frac{1}{k!}\sum_{\ell=1}^k \frac{\left|\text{det}_\mp\left[\mathfrak{S}_{\boldsymbol{\alpha i}\setminus\{\ell\}}^{\boldsymbol{\zeta z}\setminus\{k\}}\right]\right|^2}{\braket{\boldsymbol{\alpha i}\setminus\{\ell\}|\boldsymbol{\alpha i}\setminus\{\ell\}}}= \frac{1}{k} \sum_{\ell=1}^k P(\zeta_1z_1,\cdots, \zeta_{k-1}z_{k-1} |\boldsymbol{\alpha i}\setminus \{\ell\}),
\end{split}
\end{equation}
where $\boldsymbol{\alpha i}\setminus\{\ell\}$ corresponds to the ordered sequence $\alpha_1i_1\leq \cdots\leq \alpha_{\ell-1}i_{\ell-1}\leq\alpha_{\ell+1}i_{\ell+1}\leq\cdots\leq \alpha_ki_k$ obtained by removing the $\ell$-th element. Concretely, we are ``simplifying" the initial state $\boldsymbol{\alpha i}$ by removing one particle from it.
To obtain Eq.~\eqref{app:eq:probability-property}, we first used the unitarity of the scattering matrix and then summed out the Kronecker delta.

As a sanity check, we verify that $P(\zeta_1z_1,\cdots, \zeta_kz_k |\boldsymbol{\alpha i})$ is normalized:
\begin{equation}
\begin{split}    
    \sum_{\zeta_1z_1}\cdots\sum_{\zeta_kz_k}&P(\zeta_1z_1,\cdots, \zeta_kz_k |\boldsymbol{\alpha i}) =\frac{1}{k}\sum_{\zeta_1z_1}\cdots\sum_{\zeta_{k-1}z_{k-1}}\sum_{\ell_1=1}^kP(\zeta_1z_1,\cdots, \zeta_{k-1}z_{k-1} |\boldsymbol{\alpha i}\setminus\{\ell_1\})\\
    &=\frac{1}{k!}\sum_{\ell_1=1}^k\sum_{\substack{\ell_2\neq \ell_1;\\ \ell_2=1}}^k\cdots \sum_{\substack{\ell_k\neq \ell_1,\cdots,\ell_{k-1};\\\ell_k=1}}^k P(\varnothing|\boldsymbol{\alpha i}\setminus\{\ell_1\}\cdots\setminus\{\ell_k\})=1,
\end{split}
\end{equation}
where we recursively used the coarse-graining property Eq.~\eqref{app:eq:probability-property} to remove one-by-one each particle in the initial state $\boldsymbol{\alpha i}$. We are left with the probability $P(\varnothing|\varnothing)$ of observing no output particles given no input particles, which is clearly 1. 

Next, the (stochastic) number of particles in state $1x$ of the output state is
\begin{equation}
 n_{1x}^{\boldsymbol{\zeta z}} = \sum_{j=1}^k \delta_{\zeta_j z_j, 1x},
\end{equation}
which can be $0,1$ for fermions and $0,\ldots,k$ for bosons. Given the invariance under permutation of the conditional probability Eq.~\eqref{app:eq:unordered-conditional-probability}, we now focus on the first Kronecker delta. Its expectation value conditioned on the initial state reads
\begin{equation}
    \begin{split}
        \Ex{\delta_{\zeta_1z_1,1x}|\boldsymbol{\alpha i}}&=\sum_{\zeta_2z_2}\cdots\sum_{\zeta_k z_k}P(1x,\cdots,\zeta_kz_k|\boldsymbol{\alpha i})\\
        &=\frac{1}{k!}\sum_{\ell_1=1}^k\sum_{\substack{\ell_2\neq \ell_1;\\ \ell_2=1}}^k\cdots\cut \sum_{\substack{\ell_{k-1}\neq \ell_1,\cdots,\ell_{k-2};\\\ell_{k-1}=1}}^k\cut P(1x|\boldsymbol{\alpha i}\setminus\{\ell_1\}\cdots\setminus\{\ell_{k-1}\}) \\
        &= \frac1k\sum_{y=1}^k P(1x|\alpha_yi_y) = \frac1k\sum_{y=1}^k|s_{1x,\alpha_yi_y}|^2,
    \end{split}
\end{equation}
where once again we used Eq.~\eqref{app:eq:probability-property} and noticed that there are $(k-1)!$ possible ways to remove all but $\alpha_yi_y$ from the initial state. Crucially, this expectation value only depends on the probabilities of single-particle processes.
Then, the average number of transferred particles in state $1x$ is given by 
\begin{equation}
    \begin{split}
        \Ex{n^{\boldsymbol{\zeta z}}_{1x} - n^{\boldsymbol{\alpha i}}_{1x}} &= \sum_{n_{11}^{\boldsymbol{\alpha i}},\cdots,n_{LC_L}^{\boldsymbol{\alpha i}}}\left(\sum_{y=1}^kP(1x|\alpha_yi_y) - n_{1x}^{\boldsymbol{\alpha i}}\right)p_{11}(n_{11}^{\boldsymbol{\alpha i}})\cdots p_{LC_L}(n_{LC_L}^{\boldsymbol{\alpha i}})\\
        &= \sum_{\alpha i} P(1x|\alpha i)\langle n_{\alpha i}\rangle - \langle n_{1x}\rangle  = \sum_{\alpha i} |s_{1x,\alpha i}|^2(f_\alpha(\epsilon_{\alpha i}) - f_1(\epsilon_{1x})),
    \end{split}
\end{equation}
where in the last step we used the unitarity of the scattering matrix. 
To calculate the variance of the number of transferred particles we need correlators of the kind
\begin{equation}
\begin{split}
    \Ex{\delta_{\zeta_1z_1,1x}\delta_{\zeta_2z_2,1y}|\boldsymbol{\alpha i}} &=\sum_{\zeta_3z_3}\cdots\sum_{\zeta_k z_k}P(1x,1y,\cdots,\zeta_kz_k|\boldsymbol{\alpha i})\\
    &= \frac{2}{k(k-1)}\sum_{z_1< z_2}P(1x,1y|\alpha_{z_1}i_{z_1},\alpha_{z_2}i_{z_2}).
\end{split}
\end{equation}
Indeed, the conditioned expectation value of $n_{1x}^{\boldsymbol{\zeta z}}n_{1y}^{\boldsymbol{\zeta z}}$ is
\begin{equation}
\begin{split}
\Ex{n_{1x}^{\boldsymbol{\zeta z}}n_{1y}^{\boldsymbol{\zeta z}}|\boldsymbol{\alpha i}} &= \sum_{j,j'}\Ex{\delta_{\zeta_jz_j,1x}\delta_{\zeta_{j'}z_{j'},1y}} = \sum_j \Ex{\delta_{\zeta_jz_j,1x}\delta_{\zeta_{j}z_{j},1y}} + \sum_{j\neq j'}\Ex{\delta_{\zeta_jz_j,1x}\delta_{\zeta_{j'}z_{j'},1y}} \\
&= \delta_{xy}\sum_{z_1} P(1x|\alpha_{z_1}i_{z_1}) + 2 \sum_{z_1<z_2}P(1x,1y|\alpha_{z_1}i_{z_1},\alpha_{z_2}i_{z_2}).
\end{split}
\end{equation}
We use this to calculate the second moment of the transferred observable $O_1 = \sum_xo_{1x}(n_{1x}^{\boldsymbol{\zeta z}} - n_{1x}^{\boldsymbol{\alpha i}})$, where $o_{1x}$ is the generalized charge of state $1x$. For particle and energy transfer we have $o_{1x}=1$ and $o_{1x} = \epsilon_{1x}$, respectively. For the thermodynamic entropy we have $o_{1x} = \log\left[\frac{1\mp f_1(\epsilon_{1x})}{f_1(\epsilon_{1x})}\right]$.
The conditioned second moment of $O_1$ is
\begin{equation}
    \begin{split}
        &\Ex{\left[\sum_x o_{1x}(n_{1x}^{\boldsymbol{\zeta z}} - n_{1x}^{\boldsymbol{\alpha i}})\right]^2\big|\boldsymbol{\alpha i}} =\\
        &= \sum_{xy}o_{1x}o_{1y} \left\{\Ex{n_{1x}^{\boldsymbol{\zeta z}}n_{1y}^{\boldsymbol{\zeta z}}|\boldsymbol{\alpha i}} + n_{1x}^{\boldsymbol{\alpha i}}n_{1y}^{\boldsymbol{\alpha i}} - n_{1y}^{\boldsymbol{\alpha i}}\Ex{n_{1x}^{\boldsymbol{\zeta z}}|\boldsymbol{\alpha i}} - n_{1x}^{\boldsymbol{\alpha i}}\Ex{n_{1y}^{\boldsymbol{\zeta z}}|\boldsymbol{\alpha i}}\right\}\\
        &= \sum_{xy} o_{1x}o_{1y}\left\{\sum_{z=1}^k\left[(\delta_{1x,1y}- n_{1y}^{\boldsymbol{\alpha i}})P(1x|\alpha_{z}i_{z})  - n_{1x}^{\boldsymbol{\alpha i}} P(1y|\alpha_{z}i_{z})\right]+\right.\\
        &\left.\QQuad\QQuad +2\sum_{z_1<z_2}P(1x,1y|\alpha_{z_1}i_{z_1},\alpha_{z_2}i_{z_2}) + n_{1x}^{\boldsymbol{\alpha i}}n_{1y}^{\boldsymbol{\alpha i}} \right\}.
    \end{split}
\end{equation}
Taking the average over the initial state $\boldsymbol{\alpha i}$ and subtracting the squared first momen,t we find the variance
\begin{equation}\label{app:eq:Ovariance}
\begin{split}
    &\Var{\sum_x o_{1x}(n_{1x}^{\boldsymbol{\zeta z}}-n_{1x}^{\boldsymbol{\alpha i}})} = \sum_{xy}o_{1x}o_{1y}\left\{\text{Cov}[n_{1x}^{\boldsymbol{\alpha i}},n_{1y}^{\boldsymbol{\alpha i}}] + \delta_{1x,1y}\sum_{\beta j}\Ex{n_{\beta j}^{\boldsymbol{\alpha i}}}P(1x|\beta j)+ \right.\\
    & -\!\!\sum_{\beta j}\!\!\left(\text{Cov}[n_{1x}^{\boldsymbol{\alpha i}},n_{\beta j}^{\boldsymbol{\alpha i}}]P(1y|\beta j) + \text{Cov}[n_{\beta j}^{\boldsymbol{\alpha i}},n_{1y}^{\boldsymbol{\alpha i}}]P(1x|\beta j)\right) -\sum_{\beta j}\Ex{n_{\beta j}^{\boldsymbol{\alpha i}}}P(1x,1y|\beta j,\beta j) +\\
    & \left.+\!\!\cut\sum_{\beta_1j_1,\beta_2j_2}\!\!\cut\left(\Ex{n_{\beta_1 j_1}^{\boldsymbol{\alpha i}}n_{\beta_2 j_2}^{\boldsymbol{\alpha i}}}P(1x,1y|\beta_1j_1,\beta_2j_2) - \Ex{n_{\beta_1 j_1}^{\boldsymbol{\alpha i}}}\Ex{n_{\beta_2 j_2}^{\boldsymbol{\alpha i}}}P(1x|\beta_1j_1)P(1y|\beta_2j_2)\right)\right\}.
\end{split}
\end{equation}
Note that one can also calculate the covariance between the transferred observables $O_1, O_2$ in different leads. This is done in Eq.~\eqref{app:eq:Ovariance} by replacing $1y\to 2y$.
Using that different leads are initially uncorrelated and their occupation numbers are determined by the fermionic/bosonic statistics of Eq.~\eqref{eq:fermi-bose-statistics}, as well as the transmission probability of Eq.~\eqref{app:eq:unordered-conditional-probability}, we evaluate the variance of a transferred observable Eq.~\eqref{app:eq:Ovariance}.
For conciseness, we introduce the matrices
\begin{equation}\label{app:eq:matrices_definition}
\begin{split}
    &{(\boldsymbol{O}_1)}_{xy} := o_{1x}\delta_{xy},\quad {(\boldsymbol{f}_\beta)}_{jk} := f_{\beta}(\epsilon_{\beta j})\delta_{jk},\\
    &{(\boldsymbol{t}_{\alpha\beta})}_{ij}:= s_{\alpha i,\beta j},\quad\boldsymbol{X}_1 :=\boldsymbol{O}_1\left(\sum_\beta \boldsymbol{t}_{1\beta}\boldsymbol{f}_\beta\boldsymbol{t}_{1\beta}^\dagger - \boldsymbol{f}_1\right),
\end{split}
\end{equation}
such that the first and second cumulants of the transferred observable read
\begin{subequations}\label{app:eq:Oavg-var}
    \begin{align}
        \Ex{O_1} &= \Tr\left\{\boldsymbol{X}_1\right\} = \Tr\left\{\boldsymbol{O}_1\left(\sum_\beta \boldsymbol{t}_{1\beta}\boldsymbol{f}_\beta\boldsymbol{t}_{1\beta}^\dagger - \boldsymbol{f}_1\right)\right\},\\
        \Var{O_1} &= \sum_\beta \Tr\left\{\boldsymbol{O}_1^2 \left(\boldsymbol{f}_1\boldsymbol{t}_{1\beta}(1\mp\boldsymbol{f}_\beta)\boldsymbol{t}_{1\beta}^\dagger + (1\mp\boldsymbol{f}_1)\boldsymbol{t}_{1\beta}\boldsymbol{f}_\beta\boldsymbol{t}_{1\beta}^\dagger\right)\right\}+\nonumber \\
        &\qquad\mp \Tr\left\{\boldsymbol{X}_1^2\right\} -2\Tr\left\{\boldsymbol{O}_1\boldsymbol{t}_{11}\boldsymbol{O}_1 \boldsymbol{f}_1(1\mp\boldsymbol{f}_1)\boldsymbol{t}_{11}^\dagger\right\}.\label{app:eq:Ovar}
    \end{align}
\end{subequations}
When $[\boldsymbol{O}_1\boldsymbol{f}_1, \boldsymbol{t}_{11}]=0$, the variance simplifies since the $\beta =1$ contribution cancels out with the last term.
Note that in Eq.~\eqref{app:eq:Oavg-var} the trace spans also over the energy degree of freedom. These formulas can be made less compact by keeping track explicitly of the energy dependence.
We do so by specifying the indices as in the main text, see Eq.~\eqref{eq:index_replacement}. The energy dependence of the matrices in Eq.~\eqref{app:eq:matrices_definition} specifies the sub-block considered.
In particular, the matrix $\boldsymbol{t}_{1\beta}(\epsilon_{1l}, \epsilon_{\beta l'})$ accounts for the transmission amplitudes from the channels in lead $\beta$ at energy $\epsilon_{\beta l'}$ to channels in lead $1$ at energy $\epsilon_{1l}$. In periodically driven systems, this matrix is non-zero only when $\epsilon_{\beta l'} = \epsilon_{1 l} + n\hbar\Omega$, with $n\in\mathbb{Z}$, and $\Omega = 2\pi/T$ is the frequency of the drive.
Keeping track of the energy dependence in the average of the transferred observable we have
\begin{equation}
    \begin{split}
        &\Ex{O_1} = \sum_{l} \Tr\left\{\boldsymbol{O}_1(\epsilon_{1l})\left(\sum_{\beta l'}\boldsymbol{t}_{1\beta}(\epsilon_{1l}, \epsilon_{\beta l'})\boldsymbol{f}_\beta(\epsilon_{\beta l'})\boldsymbol{t}_{1\beta}^\dagger(\epsilon_{1l}, \epsilon_{\beta l'}) - \boldsymbol{f}_{1}(\epsilon_{1l})\right)\right\} \\
        &=\sum_{l} \Tr\left\{\boldsymbol{O}_1(\epsilon_{1l})\left(\sum_{\beta l'}\boldsymbol{t}_{1\beta}(\epsilon_{1l}, \epsilon_{1l}+\hbar\Omega l')\boldsymbol{f}_\beta(\epsilon_{1l}+\hbar\Omega l')\boldsymbol{t}_{1\beta}^\dagger(\epsilon_{1l}, \epsilon_{1l}+\hbar\Omega l') - \boldsymbol{f}_{1}(\epsilon_{1l})\right)\right\}\\
        &=\int dE g_1(E) \Tr\left\{\boldsymbol{O}_1(E)\left(\sum_{\beta l'}\boldsymbol{t}_{1\beta}(E, E_{l'})\boldsymbol{f}_\beta(E_{l'})\boldsymbol{t}_{1\beta}^\dagger(E, E_{l'}) - \boldsymbol{f}_{1}(E)\right)\right\},
    \end{split}
\end{equation}
where $g_1(E) = \sum_l \delta(E-\epsilon_{1l})$ is the density of states of lead 1.
Repeating the same procedure for the variance of the transferred observable we find
\begin{equation}
    \begin{split}
        &\Var{O_1}= \int dE g_1(E)\left(\Tr\Bigg\{\boldsymbol{O}_1^2(E)\sum_{\beta l}\left[\boldsymbol{f}_1(E)\boldsymbol{t}_{1\beta}(E,E_l)[1\mp \boldsymbol{f}_\beta(E_l)]\boldsymbol{t}_{1\beta}^\dagger (E,E_l)+\right.\right.\\
        &\left.+[1\mp\boldsymbol{f}_1(E)]\boldsymbol{t}_{1\beta}(E,E_l) \boldsymbol{f}_\beta(E_l)\boldsymbol{t}_{1\beta}^\dagger (E,E_l)\right] \Bigg\}+ \\
        &\mp\Tr\left\{\sum_{l'}\boldsymbol{O}_1(E)\left[\sum_{\beta l}\boldsymbol{t}_{1\beta}(E,E_l) \boldsymbol{f}_\beta(E_l)\boldsymbol{t}_{1\beta}^\dagger (E_{l'},E_l) -\boldsymbol{f}_1(E)\delta_{l',0}\right]\right.\times\\
        &\left.\times\boldsymbol{O}_1(E_{l'})\left[\sum_{\gamma q}\boldsymbol{t}_{1\gamma}(E_{l'},E_{q}) \boldsymbol{f}_\gamma(E_q)\boldsymbol{t}_{1\gamma}^\dagger (E,E_q) -\boldsymbol{f}_1(E)\delta_{l',0}\right]\right\} + \\
        &-2\left.\Tr\left\{\boldsymbol{O}_1(E)\sum_l\boldsymbol{t}_{11}(E,E_l)\boldsymbol{O}_1(E_l)\boldsymbol{f}_1(E_l)[1\mp \boldsymbol{f}_1(E_l)]\boldsymbol{t}_{11}^\dagger (E,E_l)\right\}\right).
    \end{split}
\end{equation}
Further simplifications happen when (i) the considered observable and the occupation number do not depend on the channel, i.e. $\boldsymbol{O}_1(E) = O_1(E) \mathbbm{1}_1(E), \boldsymbol{f}_\beta(E)=f_\beta(E)\mathbbm{1}_\beta(E)$, and (ii) the scattering does not change the single-particle energy, i.e. the system is stationary and $\boldsymbol{t}_{1\beta} (E,E_l)=\delta_{l0}\boldsymbol{t}_{1\beta}(E,E)$. These assumptions guarantee $[\boldsymbol{O}_1\boldsymbol{f}_1, \boldsymbol{t}_{11}]=0$, 
and allow us to write the variance as
\begin{equation}
    \begin{split}
        \Var{O_1}= \int& dE g_1\left(O_1^2\sum_{\beta\neq 1}\Tr\left\{\boldsymbol{t}_{1\beta}^\dagger \boldsymbol{t}_{1\beta}\right\}\left(f_1[1\mp {f}_\beta] + [1\mp f_1]f_\beta\right)+\right.\\
        &\left.\mp O_1^2\sum_{\beta,\gamma}[f_\beta -f_1][f_\gamma -f_1] \Tr\left\{\boldsymbol{t}_{1\beta} \boldsymbol{t}_{1\beta}^\dagger\boldsymbol{t}_{1\gamma} \boldsymbol{t}_{1\gamma}^\dagger\right\}\right),
    \end{split}
\end{equation}
where we dropped the energy argument for conciseness.
\section{Two-point measurement and full counting statistics}\label{app:FCS}
In this Appendix, we consider the statistics obtained from a general two-point measurement scheme, showing that the moment generating function associated with the accumulation of an observable $\hat{O}$ can be used to derive the expressions for the full counting statistics in scattering theory in the appropriate limit~\cite{Bachmann2008Aug,Esposito2009Dec}. Consider the two-point measurement scheme applied to the measurement of the physical observable $\hat{O} = \sum_{i}o_i\hat\Pi_{o_i}$.
Calling $o_\mathrm{i},o_\mathrm{f}$ the outcomes of the initial and final measurement respectively, the joint probability distribution reads
\begin{equation}\label{app:eq:2pm-general}
    P(o_\mathrm{f}, t; o_\mathrm{i}, 0) = \Tr\left\{\hat\Pi_{o_\mathrm{f}}U\hat\Pi_{o_\mathrm{i}}\rho_0\hat\Pi_{o_\mathrm{i}} U^\dagger \hat\Pi_{o_\mathrm{f}}\right\},
\end{equation}
where the second measurement is performed after time $t$ from the first measurement, which happens at time 0 on the initial state $\rho_0$.
The unitary transformation $U$ between the two measurements is given by
\begin{equation}
    U = \mathcal{T} \exp\left[-\frac{i}{\hbar}\int_0^t \hat{H}(s)ds\right],
\end{equation}
where $\mathcal{T}$ denotes time ordering.

A useful way to evaluate the moments of the change in the observed outcome $o_\mathrm{f}-o_\mathrm{i}$ is through the moment generating function, defined as $\mathcal{G}(\chi):=\mathbb{E}[e^{i\chi(o_\mathrm{f}-o_\mathrm{i})}]$.
Using the joint probability Eq.~\eqref{app:eq:2pm-general}, the moment generating function reads
\begin{equation}
\begin{split}
    \mathcal{G}(\chi) &= \sum_{o_\mathrm{f},o_\mathrm{i}}e^{i(o_\mathrm{f}-o_\mathrm{i})\chi}P(o_\mathrm{f},t;o_\mathrm{i},0)\\
    &=\Tr\left\{\left[\sum_{o_\mathrm{f}}e^{io_\mathrm{f}\chi}\hat\Pi_{o_\mathrm{f}}\right]U \left[\sum_{o_\mathrm{i}} e^{-io_\mathrm{i}\chi}\hat\Pi_{o_\mathrm{i}}\rho_0\hat\Pi_{o_\mathrm{i}}\right] U^\dagger\right\}\\
    &=\Tr\left\{U^\dagger e^{i\hat{O}\chi}U \left[\sum_{o_\mathrm{i}} e^{-io_\mathrm{i}\chi}\hat\Pi_{o_\mathrm{i}}\rho_0\hat\Pi_{o_\mathrm{i}}\right] \right\} =\Tr\left\{ e^{i\hat{O}_\mathrm{H}(t)\chi} \left[\sum_{o_\mathrm{i}} e^{-io_\mathrm{i}\chi}\hat\Pi_{o_\mathrm{i}}\rho_0\hat\Pi_{o_\mathrm{i}}\right] \right\}. \\
\end{split}
\end{equation}
Here, we introduced the operator $\hat{O}_\text{H}(t):= U^\dagger \hat{O} U$ in the Heisenberg picture.

When the initial state commutes with the measurement projectors $\hat\Pi_{o_\mathrm{i}}$, i.e.
\begin{equation}
    [\hat\Pi_{o_\mathrm{i}}, \rho_0]=0\quad \forall o_\mathrm{i},
\end{equation}
the initial state is not perturbed by the first measurement, and the moment generating function becomes
\begin{equation}\label{eq:2PM-generating-function}
\mathcal{G}(\chi) = \Tr\left\{e^{i\hat{O}_\mathrm{H}(t)\chi}e^{-i\hat{O}\chi}\rho_0\right\} = \Tr\left\{\mathcal{T}\left\{e^{i\hat{Q}_\mathrm{H}(t)\chi}\right\}\rho_0\right\} 
\end{equation}
where the operator $\hat{Q}_\mathrm{H}(t)$ represents the accumulation of the physical observable $\hat{O}$ in the process.
Using Heisenberg's equation of motion to introduce the current operator $\hat{J}^{(O)}_\mathrm{H}(t)$,
\begin{equation}
    \partial_t \hat{O}_\mathrm{H}(t) = i[\hat{H}_\mathrm{H}(t), \hat{O}_\mathrm{H}(t)] =: \hat{J}^{(O)}_\mathrm{H}(t),
\end{equation}
where we assumed the observable $\hat{O}$ to be time-independent in the Schr\"odinger picture, we write the accumulation $\hat{Q}_\mathrm{H}(t)$ as
\begin{equation}
    \hat{Q}_\mathrm{H}(t) =  \hat{O}_\mathrm{H}(t) - \hat{O}=\int_0^t \hat{J}^{(O)}_\mathrm{H}(s)ds.
\end{equation}

Note that the Lesovik-Levitov formula~\cite{Levitov1993} for the full counting statistics of the scattering process can be derived from Eq.~\eqref{eq:2PM-generating-function}~\cite{Klich2003,Schonhammer2007May}.
\section{Scaled cumulants and current statistics}\label{app:scaled-cumulants}

We now combine what we discussed in Appendices~\ref{app:probability-distribution} and \ref{app:FCS} to make the connection between the current and noise of Sec.~\ref{sec:scattering_theory} and those obtained in scattering theory~\cite{Blanter2000Sep, Moskalets2011Sep} concrete.
We therefore evaluate the first and second moment of the accumulated observable
\begin{subequations}
\begin{align}
    \Ex{o_\text{f}-o_\text{i}}&= -i\partial_\chi \mathcal{G}\big|_{\chi=0} = \Tr\left\{\int_0^t \hat{J}_\text{H}^{(O)}(s) ds \rho_0\right\},\\
    \Ex{(o_\text{f}-o_\text{i})^2} &= -\partial_\chi^2 \mathcal{G}\big|_{\chi=0} = \Tr\left\{\int_0^t\int_0^t \mathcal{T}\left\{\hat{J}_\text{H}^{(O)}(s)\hat{J}_\text{H}^{(O)}(s')\right\} dsds' \rho_0\right\}.
\end{align}
\end{subequations}
In the long time limit, the scaled first and second cumulants are
\begin{subequations}
    \begin{align}
    J^{(O)} &:= \lim_{t\to\infty} \frac{\Ex{o_\text{f}-o_\text{i}}}t =\lim_{t\to\infty} \frac1t \int_0^t \langle\hat{J}_\text{H}^{(O)}(s)\rangle ds,
    \\
    \S^{(O)} &:= \lim_{t\to\infty} \frac{\Var{o_\text{f}-o_\text{i}}}t = \lim_{t\to\infty} \frac2t \int_0^t ds \int_0^{t-s} ds' \langle \delta \hat{J}_\text{H}^{(O)}(s+s')\delta\hat{J}_\text{H}^{(O)}(s)\rangle,\label{app:eq:scaled-second-cumulant}
    \end{align}    
\end{subequations}
with $\delta\hat{J}_\mathrm{H}^{(O)}(s)=\hat{J}_\mathrm{H}^{(O)}(s)-\langle \hat{J}_\mathrm{H}^{(O)}(s)\rangle$.

We now consider a periodically driven system, such that the expectation values of the currents have period $T$, e.g. $\langle \hat{J}_\text{H}^{(O)}(s+T)\rangle =\langle \hat{J}_\text{H}^{(O)}(s)\rangle$.
Concretely, this is the case when the Hamiltonian is $T$-periodic, $\hat{H}(s+T) = \hat{H}(s)$, and the initial state commutes with the unitary evolution of a period $[\rho_0, U(T,0)]=0$.
Then, the scaled first cumulant becomes the period-averaged current, 
\begin{equation}
    J^{(O)} = \lim_{n\to\infty}\frac{1}{nT + \tau}\left(n\int_0^T\langle\hat{J}_\text{H}^{(O)}(s)\rangle ds + \int_0^\tau \langle \hat{J}_\text{H}^{(O)}(s)\rangle ds\right) = \frac{1}{T}\int_0^T \langle \hat{J}_\text{H}^{(O)}(s)\rangle ds.
\end{equation}
For the scaled second cumulant, we use $\langle \delta \hat{J}_\text{H}^{(O)}(s+T)\delta \hat{J}_\text{H}^{(O)}(s'+T)\rangle=\langle \delta \hat{J}_\text{H}^{(O)}(s)\delta \hat{J}_\text{H}^{(O)}(s')\rangle$ and split the integrals as follows
\begin{equation}\label{app:eq:one-sided-noise}
    \begin{split}
        \S^{(O)} &= \lim_{n\to \infty} \frac{2}{nT+\tau}\left(\int_0^{nT}\cut ds\int_0^{nT+\tau-s}\cut ds' \langle \delta \hat{J}_\text{H}^{(O)}(s+s')\delta \hat{J}_\text{H}^{(O)}(s)\rangle \right.+\\
        &\left.\qquad +\int_0^{\tau}\cut ds\int_0^{\tau-s}\cut ds' \langle \delta \hat{J}_\text{H}^{(O)}(s+s')\delta \hat{J}_\text{H}^{(O)}(s)\rangle \right)\\
        &=\lim_{n\to \infty} \frac{2}{nT+\tau}\left(\int_0^{nT}\cut ds\int_0^{nT+\tau-s}\cut ds' \langle \delta \hat{J}_\text{H}^{(O)}(s+s')\delta \hat{J}_\text{H}^{(O)}(s)\rangle\right) \\
        &=\lim_{n\to \infty} \frac{2}{nT+\tau}\int_0^{T}\cut ds\left(\sum_{j=1}^n\int_0^{jT+\tau-s}\cut ds' \langle \delta \hat{J}_\text{H}^{(O)}(s+s')\delta \hat{J}_\text{H}^{(O)}(s)\rangle\right)\\
        &= \frac{2}{T}\int_0^T\cut ds \int_0^\infty\cut ds' \langle\delta \hat{J}_\text{H}^{(O)}(s+s')\delta\hat{J}_\text{H}^{(O)}(s)\rangle.
    \end{split}
\end{equation}
The scaled second cumulant is then the period-averaged one-sided zero-frequency current noise.

From Eq.~\eqref{app:eq:scaled-second-cumulant}, we see that $\S^{(O)}\in \mathbb{R}$ by construction, so considering $\frac12(\S^{(O)}+(\S^{(O)})^*)$ leads to
\begin{equation}
    \S^{(O)} = \frac{1}{T}\int_0^T ds \int_0^\infty ds' \langle\{\delta\hat{J}_\text{H}^{(O)}(s+s'),\delta\hat{J}_\text{H}^{(O)}(s)\}\rangle,
\label{eq:one-sided_noise}
\end{equation}
where we have now recovered the familiar anti-commutator~\cite{landauV}.
Furthermore, exploiting the $T$-periodicity, we have
\begin{equation}
    \begin{split}
        \S^{(O)} &= \frac1{T}\int_0^\infty\cut ds' \int_{s'}^{T+s'}\cut ds\langle\{\delta\hat{J}_\text{H}^{(O)}(s-s'),\delta\hat{J}_\text{H}^{(O)}(s)\}\rangle \\
        &=\frac1{T}\int_0^\infty\cut ds' \left(\int_{0}^{T} + \int_{T}^{T+s'} - \int_{0}^{s'}\right)\langle\{\delta\hat{J}_\text{H}^{(O)}(s-s'),\delta\hat{J}_\text{H}^{(O)}(s)\}\rangle ds \\
        &= \frac1{T}\int_0^\infty\cut ds' \int_{0}^{T} \langle\{\delta\hat{J}_\text{H}^{(O)}(s-s'),\delta\hat{J}_\text{H}^{(O)}(s)\}\rangle ds,
    \end{split}
\end{equation}
which allows us to write the scaled second cumulant as the (typically used) period-averaged double-sided zero-frequency current noise
\begin{equation}
    \S^{(O)} = \frac{1}{2T}\int_0^T ds \int_{-\infty}^\infty ds' \langle\{\delta \hat{J}_\text{H}^{(O)}(s+s'),\delta\hat{J}_\text{H}^{(O)}(s)\}\rangle.
\end{equation}

The stationary case is recovered in the limit $T\to 0,$ where the scaled first and second cumulant reduce to
\begin{subequations}
    \begin{align}
        J^{(O)} &= \langle \hat{J}^{(O)}\rangle =\langle\hat{J}_\mathrm{H}^{(O)}(t)\rangle, \\
        \S^{(O)} &=\frac12\int_{-\infty}^\infty ds' \langle\{\delta \hat{J}_\text{H}^{(O)}(s'),\delta\hat{J}_\text{H}^{(O)}(0)\}\rangle.
    \end{align}
\end{subequations}

The accumulation of the quantity $(o_\text{f}-o_\text{i})$ we consider is made of many statistically independent contributions $\delta O_i$, one for each scattering event.
Calling $v$ the rate at which the scattering events take place, which corresponds to the rate at which electrons cross a lead, the total accumulated observable is given by
\begin{equation}
    O = \sum_{i=1}^{v t} \delta O_i
\end{equation}
for $vt=N\gg 1$.
Then, the statistical independence of the events allows us to write the expectation value and the variance of the accumulated observable $O$ and the corresponding scaled cumulants as
\begin{equation}
\left\{
\begin{array}{ll}
        \Ex{O} &= v t\Ex{\delta O_i}\\
        \Var{O} &=v t\Var{\delta O_i}
\end{array}\right.
\quad\Rightarrow\quad\left\{\begin{array}{ll}
     J^{(O)}&= v \Ex{\delta O_i} \\
     \S^{(O)}&= v \Var{\delta O_i}
\end{array}\right..
\end{equation}
Thus, using the results of Appendix~\ref{app:probability-distribution} for the average and variance of transferred observables in a single scattering event, we find the scaled first and second cumulant.

\bibliography{refs.bib}

\nolinenumbers

\end{document}